\documentclass[aps,prb,secnumarabic,amssymb,preprint,floatfix]{revtex4}
\usepackage{amsmath}
\usepackage{latexsym}
\usepackage{natbib}
\usepackage[pdftex]{graphics}
\usepackage{epsfig}
\usepackage[english]{babel}
\usepackage{xcolor}

\newcommand{\qed}{\nobreak \ifvmode \relax \else
      \ifdim\lastskip<1.5em \hskip-\lastskip
      \hskip1.5em plus0em minus0.5em \fi \nobreak
      \vrule height0.75em width0.5em depth0.25em\fi}

\def\beq{\begin{equation}}
\def\eeq{\end{equation}}
\def\beqa{\begin{eqnarray}}
\def\eeqa{\end{eqnarray}}
\def\sub#1{_{_{#1}}}

\def\n1v{||v||_{L^1_t,L^2}}
\def\n2v{||v||_{L^\infty_t, L^2}}

\def\P{\Pi}

\def\n{{\bf n}}

\begin{document}

\title{Linear and nonlinear stability of a quasi-geostrophic mixing layer subject to a uniform background shear}

\author{Luca Biancofiore$^{1}$ and Orkan M. Umurhan$^{2,3}$}
\affiliation{
(1) Department of Mechanical Engineering
Bilkent University
06800 Bilkent, Ankara,
Turkey\\
(2) National Aeronautics and Space Administration (NASA), Ames Research Center, Space Sciences Division, Moffett Field, CA 94035, USA. \\
(3) SETI Institute, Mountain View, CA 94043, USA. \\
}
\date{\today}\date{\today}%

\begin{abstract}

The aim of this work is to shed light by revisiting - through the kernel-wave (KW) perspective - the breakdown of a quasi-geostrophic (QG) mixing layer (or vortex strip/filament) in atmosphere under the influence of a background shear.
The QG mixing layer is modelled with a family of quasi-Rayleigh velocity profiles in which the potential vorticity (PV) is constant in patches.  In the KW perspective a counter-propagating Rossby wave (CRW) is created at each PV edge, i.e. the edge where a PV jump is located. 
The important parameters of our study are (i) the vorticity of the uniform shear $m$ and (ii) the Rossby deformation radius $L_d$, which indicates how far the pressure perturbations can vertically propagate. 
While an adverse shear ($m<0$) stabilizes the system, a favorable shear ($m>0$) strengthens the instability. This is due to how the background shear affects the two uncoupled CRWs by shifting the optimal phase difference towards large (small)  wavenumber when $m<0$ ($m>0$). As the QG environment is introduced a general weakening of the instability is noticed, particularly for $m>0$. This is mainly due to the reduced interaction between the two CRWs in the QG limit.
Furthermore, nonlinear pseudo-spectral simulations in the nominally infinite Reynolds number limit were conducted using as initial base flow the same quasi-Rayleigh profiles analyzed in the linear analysis. 
The growth of the mixing layer is obstructed by introducing a background shear, especially if adverse, since the vortex pairing, which is the main growth mechanism in mixing layers, is strongly hindered. Interestingly the most energetic configuration is for $m=0$ which differs from the linear analyses for which the largest growth rates were found for a positive $m$. In absence of a background shear additional modes are subharmonically triggered by the initial disturbance enhancing the turbulent character of the flow.  We also confirm energy spectrum trends for broken-down mixing layers reported in the literature. We interpret the character of mixing-layer breakdown as being a phenomenological hybrid of Kraichnan's (Phys. Fluids, 10, 1417-1423, 1967) direct enstrophy cascade picture and the picture of self-similar vortex production.

\end{abstract}


\maketitle

\section{Introduction}


Idealized two dimensional (2D) flow and turbulence continue to serve as an indispensable platform to study the physical properties of various fluid dynamical processes and effects.  Two dimensionality arises naturally in systems exhibiting dynamically restricted dimensions like in planetary atmospheric flows and soap bubbles.  Despite often cited criticisms implicating their lack of real-world relevance, the study of idealized flows like 2D Navier-Stokes equations offers a challenging, yet intellectually traversable, setting to examine complex physical notions like, for example, the recently uncovered conformal invariance between the inverse cascade of 2D turbulence \cite{kraichnan1967inertial,Kraichnan_Montgomery_1980}
and critical phenomena of 2D statistical mechanics \cite{Bernard_etal_2006,Boffetta_Ecke_2012}.
\par
\begin{figure}
\begin{tabular}{c}
\scalebox{0.35}{\includegraphics{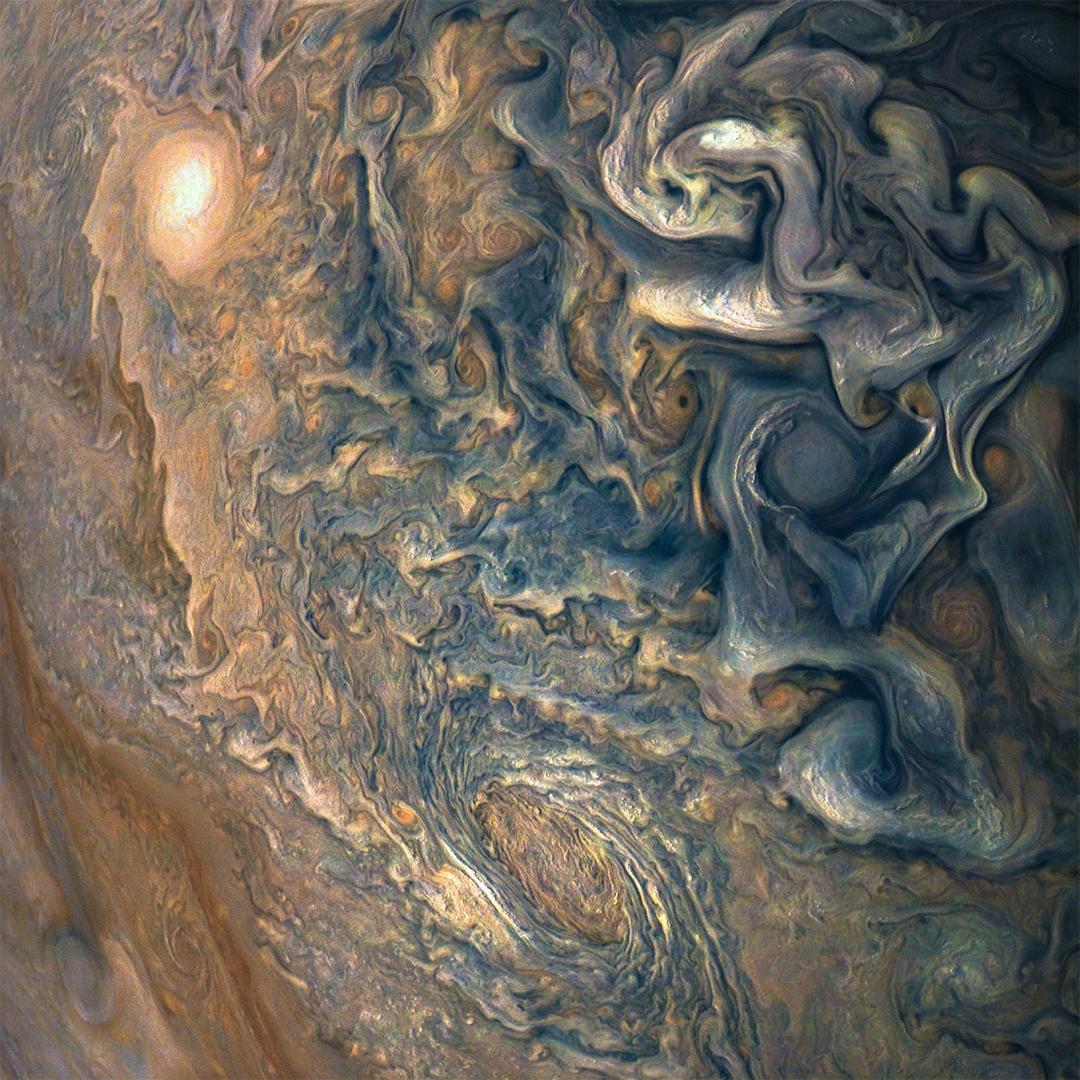}}
\end{tabular}
\caption{Juno image of Jupiter's cloud tops near northern latitude 48.9 degrees.  Horizontal scale of image is approximately 95,000 km (9.3 km/pixel).  Being in the quasigeostrophic regime, these regions exhibit filamentary structure as well as large-scale coherent vortices.  Rossby deformation radii at these latitudes is approximately 20,000 km.
Image credit: NASA/JPL-Caltech/SwRI/MSSS/Gerald Eichstadt/Sean Doran.}\label{Juno_Jupiter}
\end{figure}

Of interest to this study is the dynamical nature of the transition of vorticity strips in 2D flows. Sometimes called the {\emph {mixing layer}}, such parallel strips are idealized as streamwise oriented patches of two different {\emph{constant}} vorticities -- one value inside and another outside.  The mixing layer is considered a canonical testbed to study the breakdown of small scale filaments due to its own vorticity induced velocities with or without the influence of an externally imposed shear \cite{dritschel1989stabilization}.  This has applications to understanding the dynamics in the forward cascade inertial range of 2D turbulence studies.
The simplest mixing layer is the so-called Rayleigh layer \cite{rayleigh1945theory,drazin2012introduction,regev2016modern} in which the outside vorticity is zero which leads to a configuration in which the streamwise velocities on the outside are two different constants and the region within the strip has a streamwise velocity that linearly connects to the outer regions.  The Rayleigh layer is unstable, resulting in its roll-up into smaller coherent vortices with attendant interwoven fine scale filamentary structures. This basic mixing layer setting, as a fundamental qualitative physical model, is encountered frequently in terrestrial and planetary atmospheric flows, e.g., as recent Juno mission images of cloud top formations in Jupiter's upper atmosphere clearly shows (see figure \ref{Juno_Jupiter}).

We are partial to the counter-propagating Rossby wave (CRW) perspective \cite{bretherton1966baroclinic,hoskins1985use} in understanding the breakdown of the mixing layer \cite{heifetz1999counter,Carpenter_etal_2012}.
In this picture, first developed by Baines \& Mitsudera \cite{Baines_Mitsudera_1994}, each perturbed edge of the idealized mixing layer supports a Rossby edgewave which azimuthally propagates against the local flow.  Each Rossby edgewave instantly induces a far-field velocity (``action at a distance") so that the two edgewaves can constructively interact with one another if the conditions are right: If the sum of the intrinsic Rossby edgewave speed and the local flow velocity are nearly equal for both edgewaves then the system can enter a resonance due to the action-at-a-distance effect resulting in mutual amplitude growth.  As a general mechanism, sometimes referred to as the kernal-wave (KW) perspective\cite{Harnik_etal_2008}, this can apply to any two pairs of localized edgewaves of a mixing layer irrespective of the fundamental disturbance type -- be it gravity waves, MHD waves, capillary waves, and others \cite{sakai1989rossby, Harnik_etal_2008,rabinovich2011vorticity, heifetz2015interacting,biancofiore2015interaction,biancofiore2017understanding}. Indeed this conceptual framework has been usefully applied toward the interpretation of various phenomena like, e.g., the Holmboe instability and  Saturn's observed polar polygons \cite{umurhan2007holmboe,Ron_etal_2017}. \par

In the context of quasigeostrophic (QG) flows, typifying mid-to-high latitude synoptic scale dynamics in atmospheres and oceans \cite{Vallis_book_2006} (also see sketch in figure \ref{sketches}a), the fate of mixing layers depends sensitively on the Rossby radius of deformation \cite{waugh1991stability}, $L_d$, which is approximately 1000 km in the Earth's midlatitude atmosphere and defines the synoptic scale. On Jupiter and Saturn this figure is 15-35 times larger \cite{liu_schneider_2015}.  A cursory glance at the operator relating the streamfunction to the potential vorticity (PV) \cite{waugh1991stability}
 readily indicates that  $L_d$ acts to diminish the ability of a local patch of PV to induce far field velocity fluctuations:
large values of $L_d$ means that the stratification is strong and, consequently, the dynamical reach of a deformed edgewave is far out (and limits to the classical 2D case as $L_d\rightarrow \infty$), while when $L_d$ is small the dynamical influence of the same edgewave is muted by an exponential factor $\sim \exp \left(-r/L_d\right)$, where $r$ is the distance from the edgewave to the point of influence.
\par
Another factor characterizing a mixing layer's stability is the sense of any applied background constant-shear flow profile. 
To give an example, a uniform shear strongly affects the stability of filaments created in the vicinity of intense rotating
coherent vortices. The filaments are
rapidly aligned with the circulating flow due to the differential rotation, and then shear is the prime factor affecting
stability. If adverse, constant shear acts to suppress instability \cite{dritschel1989stabilization}.  From the viewpoint of the KW perspective, one might expect this to be so because the resonance criterion between the two opposing edgewaves is hindered as the mean-flow at the respective edges is unable to compensate for the intrinsic Rossby wave speed of the respective disturbed edges.
On its own right, examining the fate of a QG mixing layer subject to an external constant shear offers physical insight into the workings of the forward enstrophy cascade in a fully turbulent 2D setting. In particular, it sheds light upon the influence of (spectrally) non-local interactions: here being the forcing brought down from the global shear upon small scale filaments.  In this way we see the KW perspective as offering another helpful tool in understanding the characteristic development of 2D turbulence.
 \par
In this study we revisit the problem of the breakdown of mixing layers subject to an external constant shear in the QG model setting (section 2).  Our interest is to interpret the results found by Dritschel and co-workers \cite{dritschel1989stabilization,waugh1991stability} within the KW framework.  In particular, we consider the conditions leading to the mixing layer breakdown and rationalize these linear stability conditions in terms of the ability (or lack-thereof) of individual edgewaves to achieve frequency resonance as a function of the background imposed constant shear and $L_d$ (section 3).  We further test the robustness of the KW perspective by conducting a series of fully nonlinear simulations of the breakdown of QG mixing layers (section \ref{NLS}).  In particular, we recast the original setup of Dritschel and co-workers instead as a composite flow made up of a so-called quasi-Rayleigh profile plus an externally imposed steady shear and making sure -- according to the KW perspective -- that the velocities and jumps in PV are equivalent to the setup examined by Waugh \& Dritschel \cite{waugh1991stability} (section \ref{numericalsetup}). In the following section (section \ref{validation}) we validate the predictive power of the KW framework by establishing the equivalence between the measured linear growth rates of these forced simulations against those predicted by linear theory.  The rest of section \ref{NLS} is devoted to assessing the nonlinear development of the breakdown of the mixing layer in which we recover the $k^{-11/3}$ spectral energy distribution previously shown to hold in the enstrophy cascade regime of $L_d = \infty$ mixing layers \cite{gilbert1988spiral,lesieur1988mixing}. 
Furthermore, we examine the spectral energy slope in the enstrophy cascade regime as a function of $L_d$ finding that its slope ($k^{-\delta}$, $\delta$ is the slope) indeed steepens ($5/3<\delta<11/3$) as $L_d$ approaches $1$. 
We also characterize the shearwise spread of the disrupted mixing layer as a function of amplitude and sense of the applied background constant shear.  Finally, through the lens of the KW perspective, we offer an interpretation to explain the statistical quality of the broken-down shear layer as being phenomenologically indicative of a process that lies between the Kraichnan/Gilbert \cite{kraichnan1967inertial,gilbert1988spiral} explanation of enstrophy cascade in forced-dissipative 2D turbulence and the picture  of the self-similar production of coherent vortices down to the dissipative scales \cite{Santangelo_etal_1989,dritschel_1995,dritschel2008unifying}. Section \ref{conclusions} summarizes our results.

\section{Model and governing equations}

In this section we present first the model analyzed in this paper (section \ref{justmodel}). Afterwards in section \ref{KWperspective} we use the Kernel Wave (KW) perspective to arrive to a dynamical system able to describe the linear stability of this model.

\subsection{Model}\label{justmodel}

\begin{figure}
\begin{tabular}{c}
\hskip-0.2cm\scalebox{0.22}{\includegraphics{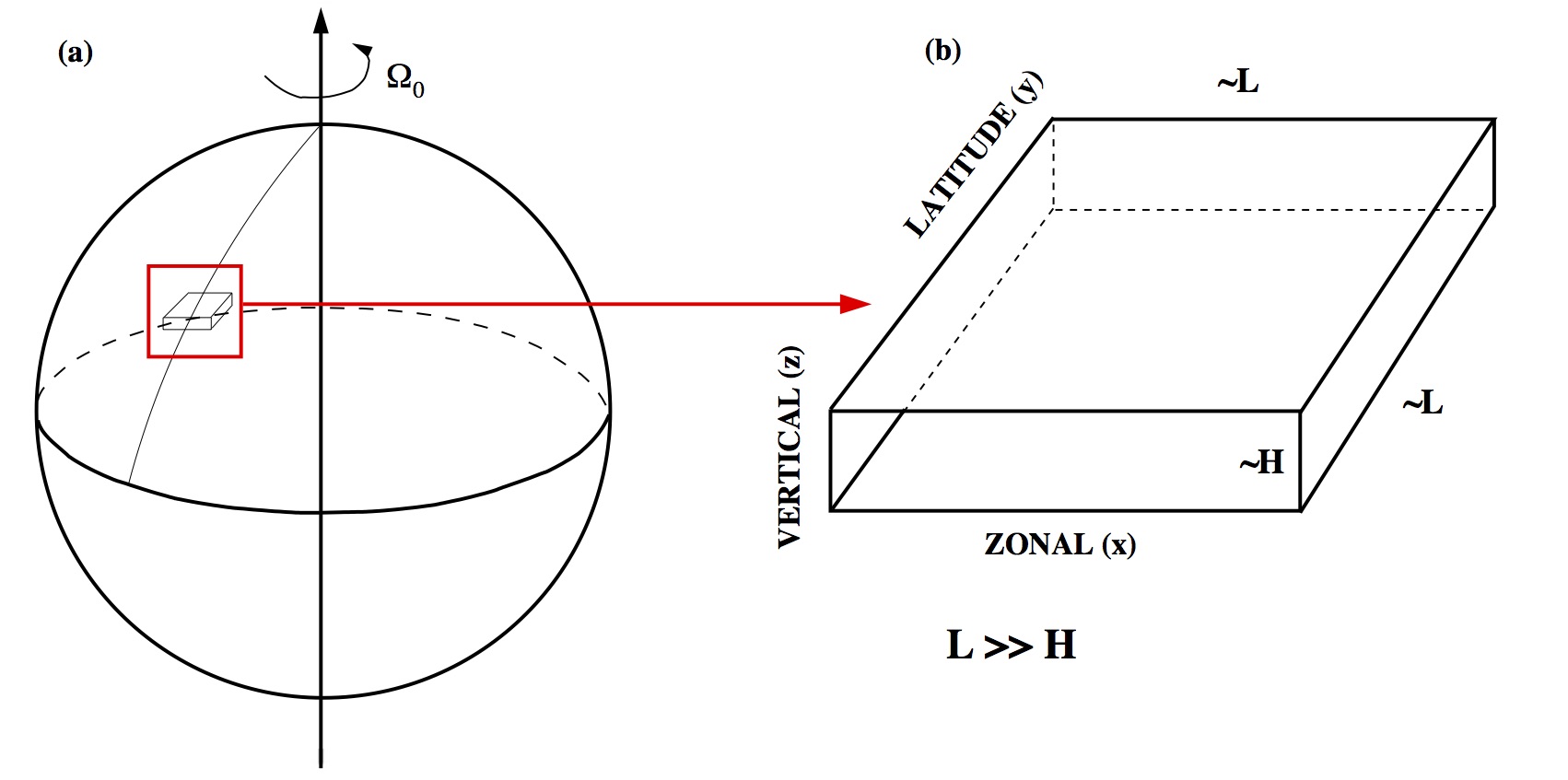}}\\
\hskip-1cm\scalebox{0.17}{\includegraphics{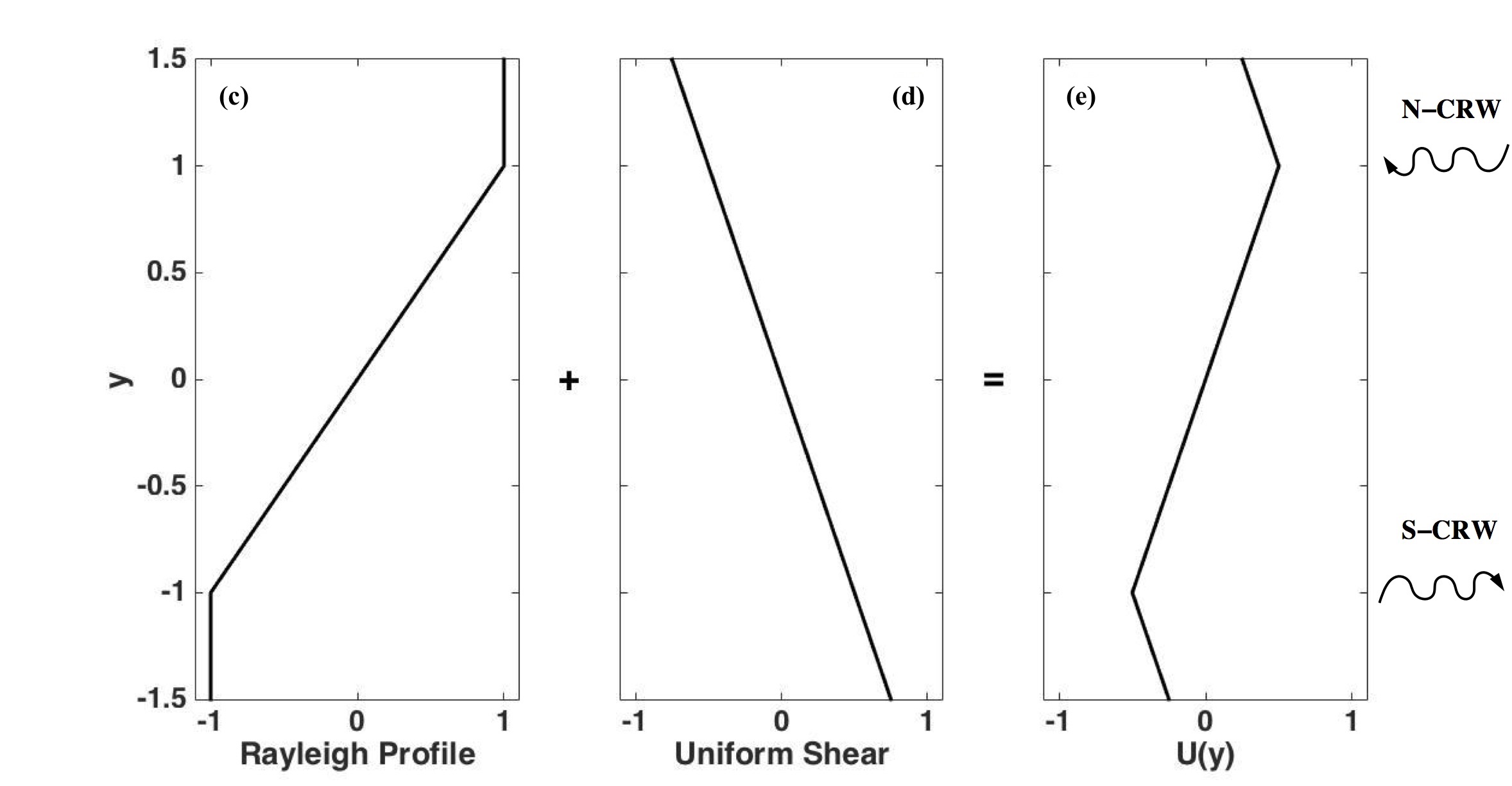}}
\end{tabular}
\caption{From (a) the planetary scale in spherical coordinates to the synoptic scale in Cartesian coordinate. (e) The velocity profile for $L_d=\infty$, i.e. (c) the Rayleigh profile plus the (d) uniform shear. Two CRWs are formed at the two vorticity edges: N-CRW at the northern edge and S-CRW at the southern edge.}\label{sketches}
\end{figure}

We consider the planetary atmosphere to be a thin fluid layer lying atop a spherical surface rotating with an angular velocity $\Omega_0$ (see fig. \ref{sketches}a). If we assume that the planetary atmosphere is a fast rotator we can consider only a Cartesian section (fig. \ref{sketches}b) where $x$ is the zonal (eastward) direction, $y$ is the latitudinal (poleward) direction and $z$ is the vertical (altitude) direction. Let $\mathcal{U}$ be the horizontal velocity scale which occurs on the horizontal length scale $L$ and let $H$ denote the vertical scale, i.e. the pressure scale-height of the atmosphere.
If we assume infinitely fast hydrostatic adjustment and to work in a $f$-plane, (i.e. no variations due to the Coriolis effect are considered in all the motions), the potential vorticity and its evolution in a two-dimensional incompressible, geostrophic flow is given by
\begin{equation}\label{potentialvorticity}
q=\nabla^2\psi-\frac{1}{L_d^2}\psi+2\Omega_0,
\qquad \frac{dq}{dt} = 0,
\end{equation}
where $\psi$ is the streamfunction given by $u=-\frac{\partial\psi}{\partial y}$ and $v=\frac{\partial\psi}{\partial x}$, where $u$ and $v$ are the zonal and latitudinal components of the velocity, respectively, $L_d$ is the Rossby radius of deformation. The Rossby radius of deformation is a measure of how far the pressure perturbations can travel at the flow time scale and is given by $L_d^2=\frac{H^2N^2}{4\Omega_0^2}$, where $N^2$ is the Brunt-V\"ais\"al\"a frequency, i.e. the frequency of the buoyancy vertical oscillations. For further details on the physically motivated derivation of eq. \ref{potentialvorticity} the reader is referred to Ref. \onlinecite{regev2016modern}.

Our model is represented by the family of the non-dimensional velocity profile already analyzed in Waugh \& Dritschel \cite{waugh1991stability}
\begin{equation}
U(y)=\left\{ \begin{array}{lll}L_d\left[e^{-\frac{y}{L_d}}\sinh(\frac{1}{L_d})+me^{-\frac{1}{L_d}}\sinh(\frac{y}{L_d})\right]&{\rm for} &
y>1\\ L_d(m+1)e^{-\frac{1}{L_d}}\sinh(\frac{y}{L_d}) &{\rm for}&
|y|< 1 \\
L_d\left[-e^{\frac{y}{L_d}}\sinh(\frac{1}{L_d})+me^{-\frac{1}{L_d}}\sinh(\frac{y}{L_d})\right]&{\rm for} &
y<-1
\end{array}\right.\label{basicvelocity1}
\end{equation}
that has piecewise constant potential vorticity 
\begin{equation}
Q(y)=\left\{ \begin{array}{lll}-me^{\frac{D-2}{2L_d}}&{\rm for} &
|y|>1\\ -\left(me^{\frac{D-2}{2L_d}}+1\right) &{\rm for}&
|y|< 1
\end{array}\right.\label{basicvorticity1}
\end{equation}
where $D$ is a distance much greater
than both the strip width (equal to $2$) and the radius of deformation $L_d$. The limit for $L_d\rightarrow\infty$ of the family velocity profile depicted by eq. \ref{basicvelocity1} (illustrated in fig \ref{sketches}e) is the Rayleigh model \cite{rayleigh1945theory,heifetz2005relating} (fig. \ref{sketches}c) sheared by a uniform shear $m$ (fig. \ref{sketches}d)
\begin{equation}
U(y)=\left\{ \begin{array}{lll}my+1&{\rm for} &
y>1\\ (m+1)y &{\rm for}&
|y|< 1 \\
my-1&{\rm for} &
y<-1.
\end{array}\right.\label{basicvelocity2}
\end{equation}

\subsection{Kernel wave perspective}\label{KWperspective}

The velocity $\textbf{v}$ and the potential vorticity $q$ can be linearized with respect to the basic state: $\textbf{v}=(U+u',v')$ and $q=Q+q'$,
where the capital letters and the primes indicate the basic state (eqs. \ref{basicvelocity1} and \ref{basicvorticity1}) and perturbation, respectively.
The linearized potential vorticity equation $\frac{Dq}{Dt}=0$ gives in a quasi-geostrophic (QG) approximation:
\begin{equation}
\left[\frac{\partial}{\partial t}+U(y)\frac{\partial}{\partial
x}\right]q'=-v'\frac{d Q(y)}{d
y}=-v'[\delta(y-1)-\delta(y+1)],\label{vorticitya}
\end{equation}
where $\delta$ symbolizes the Dirac delta function. 

Applying the KW perspective \cite{heifetz1999counter},
we look for a vorticity perturbation field that is concentrated on the discontinuities of the base flow vorticity, {\it i.e.} on the two edges of the velocity profile. Then for a single Fourier component with wavenumber $k$ of the form $e^{ikx}$, eq. \ref{vorticitya} can be rewritten as:
\begin{equation}
\hat{q}'=q_{S}(k,t)\delta\big[y+1]+q_{N}(k,t)\delta\big[y-1],\label{q}
\end{equation}
where $q_S(k,t)$ and $q_N(k,t)$ represent vorticity waves, i.e. the counterpropagating Rossby waves (CRWs), at southern and northern edge, respectively.
From eq. \ref{potentialvorticity} it follows that $\hat{q}'=-\mathcal{L}\hat{\psi}'$, with $\mathcal{L}=[d^2/dy^2-\bar{k}^2]$, where $\bar{k}^2=k^2+\frac{1}{L_d^2}$. The Green function of $-\mathcal{L}$ for an unbounded domain is
\begin{equation}
G(y,y',\bar{k})=-\frac{i}{2}\exp{(-\bar{k}|y-y'|)},\label{greenfunct}
\end{equation}
and the perturbation streamfunction is then obtained
\begin{equation}
\hat\psi'=-\frac{1}{2\bar{k}}[q_S(k,t)e^{-\bar k|y+1|}+q_N(k,t)e^{-\bar{k}|y-1|},\label{psi}
\end{equation}
where $\psi'$ satisfies $u'=\frac{\partial \psi'}{\partial
y}$, $v'=-\frac{\partial \psi'}{\partial x}$, i.e. in Fourier space $\hat{u}'=\partial_y\hat{\psi}'$ and $\hat{v}'={\rm i}k\hat{\psi}'$. Replacing equations (\ref{q}) and (\ref{psi}) into eq.
(\ref{vorticitya}), one is left with
\begin{eqnarray}\nonumber
\dot{q_{S}}\delta(y+1)+\dot{q_{N}}\delta(y-1)+{\rm i}kU(y)[q_{S}\delta(y+1)
+{q_{N}}\delta(y-1)]=\\={\rm i}\frac{k}{2\bar k}[q_S(k,t)e^{-\bar{k}|y+1|}+q_N(k,t)e^{-\bar{k}|y-1|}][\delta(y-1)-\delta(y+1)],\label{intermezzo}
\end{eqnarray}
where the notation $\dot{q}_{S,N}$ denotes $\frac{dq_{S,N}}{dt}$.

Calculating eq. (\ref{intermezzo}) at the edges, i.e. $y=\mp1$, the evolution of the vorticity perturbation is obtained:
\begin{eqnarray}\nonumber
&\dot{\pmb q}={\pmb M}\pmb q, \quad {\rm where}  \\ &\pmb M={\rm i} k\left[\begin{array}{cc}-\frac{1}{2 \bar k}+L_d(m+1)e^{-\frac{1}{L_d}}\sinh{\frac{1}{L_d}}&-\frac{e^{-2\bar k}}{2 \bar k}\\\frac{e^{-2\bar k}}{2 \bar k}&\frac{1}{2 \bar k}-L_d(m+1)e^{-\frac{1}{L_d}}\sinh{\frac{1}{L_d}}\end{array}\right].\label{qpiccolokbar}
\end{eqnarray}
The terms in the entry diagonal are the phase speed of the CRWs taken in isolation $c_{N,S}=\pm\left[\frac{1}{2 \bar k}-L_d(m+1)e^{-\frac{1}{L_d}}\sinh{\frac{1}{L_d}}\right]$, while the off-diagonal terms represent the interaction coefficient $\gamma=\frac{ke^{-2\bar k}}{2\bar k}$. The eigenvalues of the matrix $M$
\begin{equation}\label{eigenvalues}
\lambda_{1,2}=\pm\frac{1}{2}\frac{k}{\bar k}\sqrt{e^{-4\bar k}-\left[1-\bar kL_d\left(1-e^{-\frac{2}{L_d}}\right)(m+1)\right]^2}
\end{equation}
are the normal modes of the system. We note that in classical textbook treatments -- e.g., Section 6.2.4 of Vallis\cite{Vallis_book_2006} -- the above system is a 4x4 matrix system where the two extra modes correspond the responses relating to imposed boundaries (e.g., no normal flow) at a pair of finite far field positions in $y$.  Our conditions, which are appropriate for $|y|\rightarrow \infty$, are built into our modal solutions, and is the reason why our system ${\pmb M}$ is a 2x2 matrix.

\section{Linear stability analysis}\label{LSA}

In this section the linear stability of the dynamical system depicted by eq. \ref{qpiccolokbar} is analyzed. 
Particularly we study the effect of (i) the background shear in section \ref{backgroundshear} and (ii) the Rossby deformation radius in section \ref{deformradius}. Finally in section \ref{amplitudesphases} a detailed analysis about the waves amplitude and their phases is conducted.

\subsection{Influence of the uniform background shear}\label{backgroundshear}

\begin{figure}
\begin{tabular}{cc}
\hskip-0.2cm\scalebox{0.29}{\includegraphics{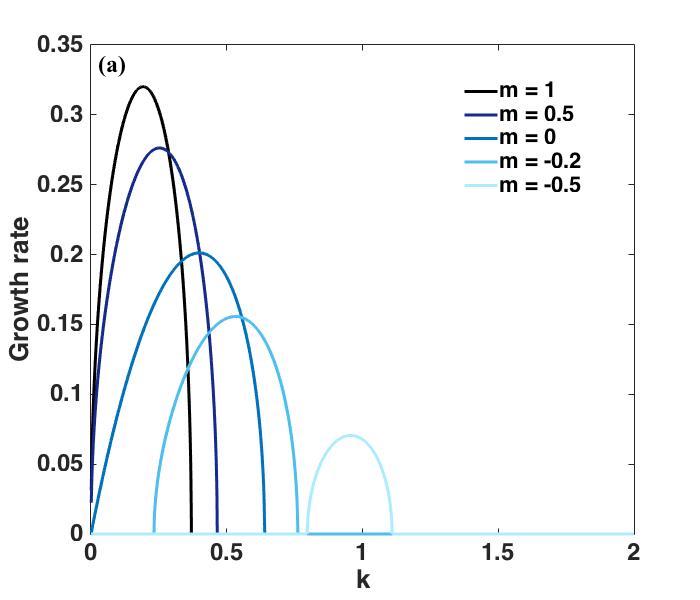}}&\hskip-0.5cm\scalebox{0.29}{\includegraphics{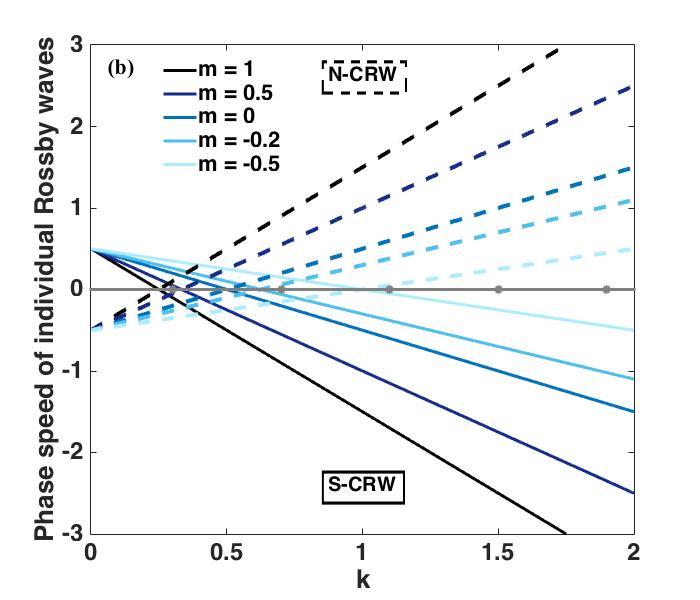}}
\end{tabular}
\caption{(a) Growth rate vs wavenumber $k$ and (b) the phase speed of the two CRWs taken in isolation for different values of the uniform shear $m$ and for $L_d=\infty$. In (b) dashed lines depict the N-CRW, while continuous lines depict S-CRW. A N-CRW (S-CRW) is counterpropagating if its phase speed is below (above) the gray line, which depicts the average base flow velocity ($U_{av}=0$).}\label{growthratem}
\end{figure}

We illustrate the growth rate $\lambda_r$, i.e. the real part of the eigenvalues in eq. \ref{eigenvalues}, in function of the wavenumber $k$ in figure \ref{growthratem}a for different values of the uniform shear background $m$ while settling $L_d=\infty$. Clearly an adverse shear ($m<0$) decreases the maximum growth rate and reduces the range of unstable wavenumbers. Moreover, this range is shifted towards small wavelengths. A positive value of $m$ conversely enhances the maximum growth rate while still decreasing the cut-off number. These results are in agreement with previous works on the effect of a background shear on the destabilization of a vorticity filament \cite{dritschel1989stabilization,waugh1991stability}.

In figure \ref{growthratem}b the phase speeds of the two CRWs taken in isolation are illustrated for the same cases of fig. \ref{growthratem}a. Dashed lines depict the northern wave (N-CRW) while continuous lines depict the southern wave (S-CRW). The range of unstable wavenumbers encloses the wavenumber $k_{equal}$ in which the individual waves have the same phase speed ($c_N=c_S$). This is consistent with the Hayashi \& Young's criterion\cite{hayashi1987stable}, i.e two waves must have a similar phase speed to be able to phase-lock and so generate the instability. 
Note that the effect of each CRW has on the other's phase speed is not considering while calculating the phase speeds (see section \ref{amplitudesphases} for more details). However this effect cannot significantly modify the phase speed for this reason the range of unstable wavenumber surrounds $k_{equal}$. Furthermore $k_{equal}$ is shifted towards smaller (large) wavenumbers for positive (negative) values of $m$. The background shear modifies the phase speeds of the CRWs allowing the interaction in a different range of wavenumbers. This explains why the range of unstable wavenumber is strongly modified by the uniform shear.

\subsection{Influence of the Rossby deformation radius}\label{deformradius}

\begin{figure}
\begin{tabular}{cc}
\hskip-0.2cm\scalebox{0.29}{\includegraphics{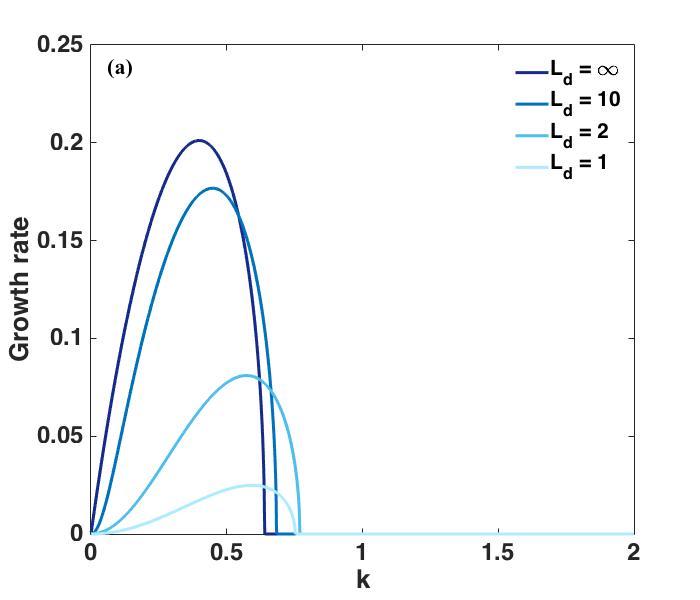}}&\hskip-0.5cm\scalebox{0.29}{\includegraphics{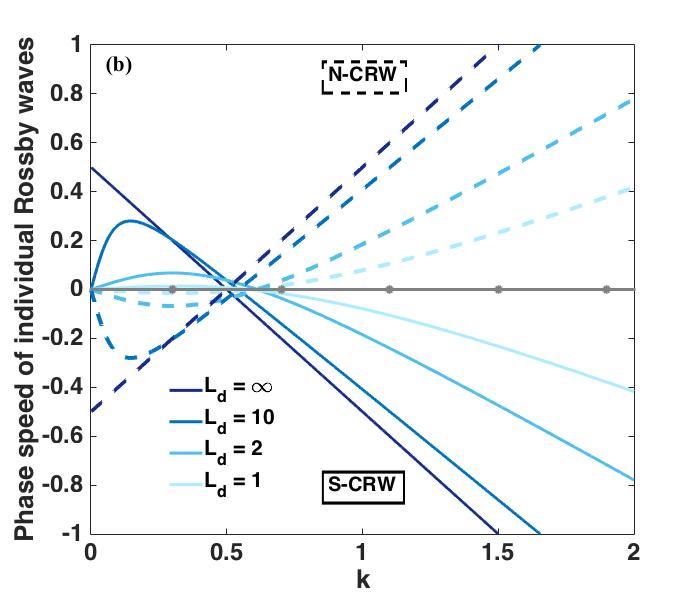}}\\
\end{tabular}
\caption{(a) Growth rate vs wavenumber $k$ and (b) the phase speed of the two CRWs taken in isolation for $m=0$ and different values of $L_d$. In (b) dashed lines depict the N-CRW, while continuous lines depict S-CRW. A N-CRW (S-CRW) is counterpropagating if its phase speed is below (above) the gray line, which depicts the average base flow velocity ($U_{av}=0$).}\label{growthrateLd} 
\end{figure}

In this section a finite value of the Rossby deformation radius is introduced. The growth rates  are illustrated in figure \ref{growthrateLd}a where the background shear is settled to $m=0$ and $L_d$ varies. The maximum growth rate is significantly damped with introducing quasi-geostrophic effects. 
A stabilization is not surprising since the reach of the CRWs is reduced when $L_d^2$ is finite. We 
rationalize this the following way: The streamfunction $\psi$ can be viewed as an equipotential surface
in the QG-context.  Taking $\Omega_0$ to be a constant,  multiplying Eq. \ref{potentialvorticity} by $\psi$ and integrating over the whole domain while making simultaneous
use of the incompressibility condition shows that 
\begin{equation}
\int_{\cal{D}}\left[ \frac{1}{2}\left(\frac{\partial \psi}{\partial x}\right)^2 
+ \frac{1}{2}\left(\frac{\partial \psi}{\partial y}\right)^2 + \frac{1}{2}\frac{\psi^2}{L_d^2}
\right]
dxdy = {\rm constant},
\end{equation}
where ${\cal D}$ is the two-dimensional whole domain.  The first two terms of the integral
expression represent the kinetic
energy contained in horizontal motions while the last containing $L_d^2$ represents a potential energy term associated with vertical displacements of the fluid layer.  From
this formulation we can readily understand that the dynamics contained in this model system
 as being distributed between horizontal velocity inducing motions and motions that
drive a local thickening of the equipotential surface.  Examining the definition of $L_d^2$, we 
see that if the Brunt-V\"ais\"al\"a frequency is weak compared to the rotational frequency, then $L_d^2$ is relatively
small which implies that the stratification, while stable, is relatively weak permitting
for longer period and larger amplitude variations of the equipotential $\psi$. 
Thus, a smaller value of $L_d^2$ means
that the potential energy term can effectively redirect and store the kinetic energy of horizontal motions.  
By storing energy, it reduces the ability of a PV disturbance to induce a far field horizontal velocity response.
Similarly, when the Brunt-V\"ais\"al\"a frequency is much more rapid compared to the rotational frequency, the fluid's relatively strong stratification means that oscillations of the
equipotential surface are rapid implying that the potential energy term is less effective at storing kinetic energy which, consequently, implies that PV disturbances have a larger reach across the domain.
The extreme end-case, $L_d^2 = \infty$, is effectively that of a finite constant density fluid layer with impenetrable horizontal walls along the vertical direction which identically permits no variation in the fluid layer's thickness -- which is also like having an infinite value of $g$.
From this point of view, a finite value of $L_d^2$ means that some horizontal flow energy can get stored as potential energy in the form of vertical (altitude) displacements of the equipotential surface and, moreover, this energy would no longer be available for non-local Rossby wave interactions. Interestingly, we note that the cut-off wavenumber does not diminish but slightly increases with decreasing $L_d^2$. 

In figure \ref{growthrateLd}b we show the phase speed of the two individual CRWs for the cases analyzed in fig. \ref{growthrateLd}a. Dashed lines represent the N-CRW, while continuous lines represent S-CRW as in fig. \ref{growthratem}b. Interestingly a decrease in the Rossby radius of deformation damps the counterpropagating nature of the waves. Less - or even no - counterpropagation is a stabilizing symptom since counterpropagation helps the phase-locking of the two waves \cite{hayashi1987stable,biancofiore2012counterpropagating}, allowing their resonance and then generating the instability. When QG effects are taken into account some of the wave energy 
associated with the Rossby wave of a given edge
is redirected into vertical storage. This stored energy would otherwise have been harnessed into inducing a velocity field at the opposite edge which would have played a role in countering the base velocity of the opposite edge. For this reason counterpropagation is thus restrained or, if the conditions are right, completely eliminated.

\begin{figure}
\begin{tabular}{cc}
\hskip-0.2cm\scalebox{0.29}{\includegraphics{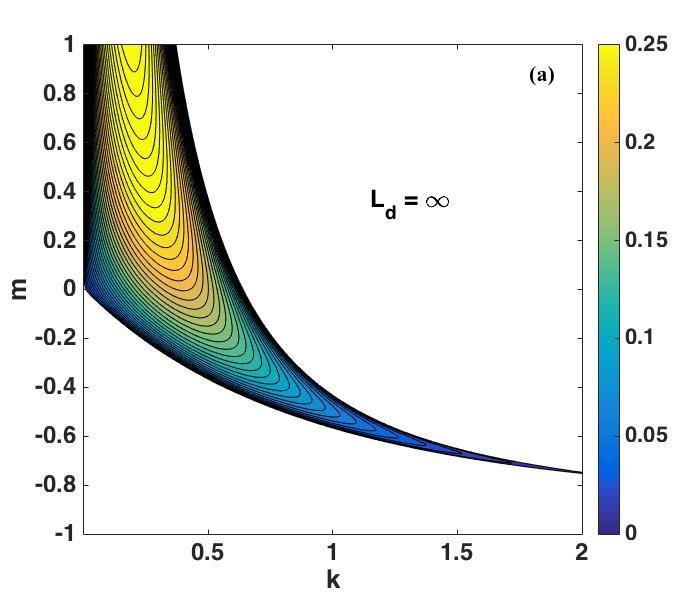}}&\hskip-0.5cm\scalebox{0.29}{\includegraphics{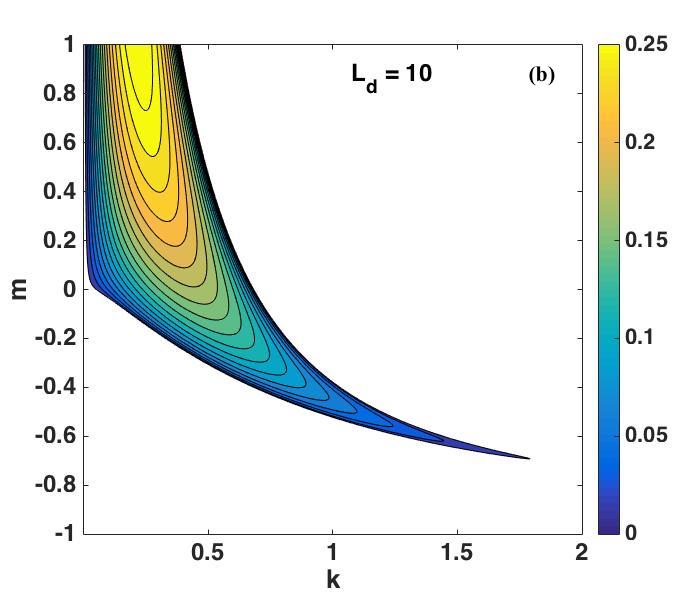}}\\
\hskip-0.2cm\scalebox{0.29}{\includegraphics{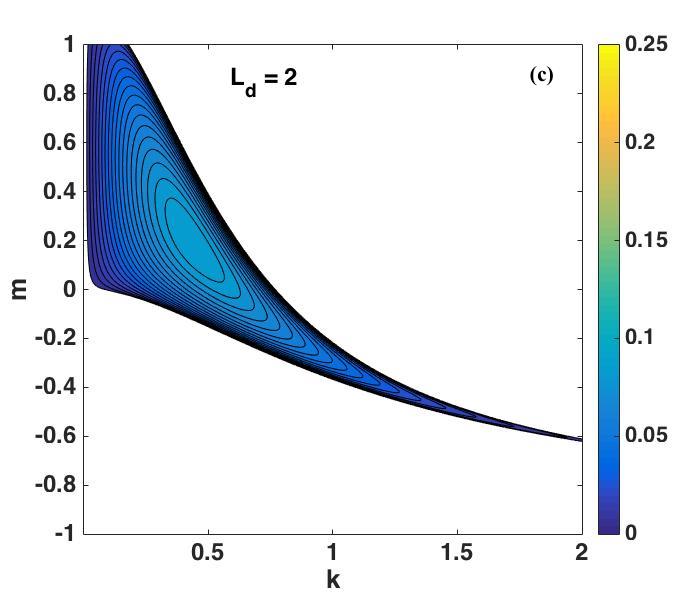}}&\hskip-0.5cm\scalebox{0.29}{\includegraphics{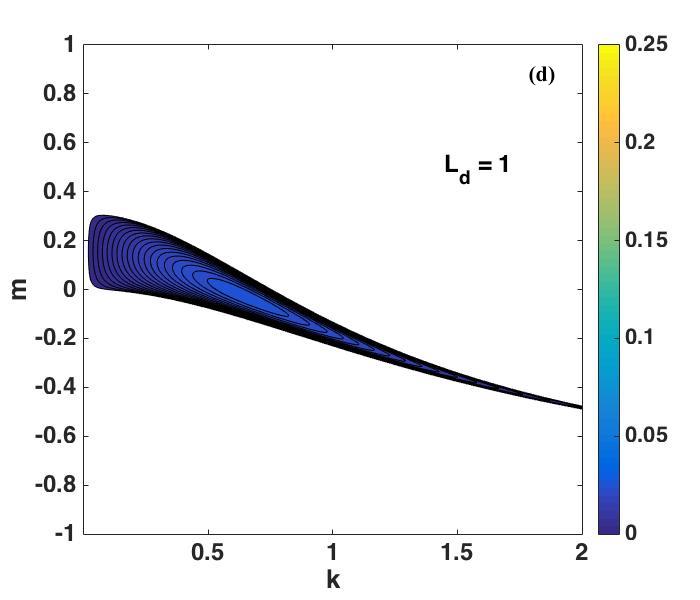}}
\end{tabular}
\caption{Growth rate in the plane $k$-$m$ for (a) $L_d=\infty$, (b) $L_d=10$, (c) $L_d=2$ and (d) $L_d=1$. 
}\label{plankmLd}
\end{figure}

In figure \ref{plankmLd} the contours of the growth rate are illustrated in the plane $k$-$m$ for (a) $L_d=\infty$, (b) $L_d=10$, (c) $L_d=2$ and (d) $L_d=1$. This figure confirms the trend previously depicted: a negative $m$ and a decrease in the Rossby deformation radius stabilize the system. In particular QG effects are more efficient for positive values of $m$.

\subsection{Amplitudes and phases of the waves}\label{amplitudesphases}

It is possible to write the two vorticity waves $q_S$ and $q_N$ in this manner
\begin{subequations}
\begin{align}
&q_S=Q_S(k,t)e^{i\epsilon_S(k,t)}\\
&q_N=Q_N(k,t)e^{i\epsilon_N(k,t)}
\end{align}
\end{subequations}
where $Q_S$ and $Q_N$ are the amplitudes of the southern/northern CRW and $\epsilon_S$ and $\epsilon_N$ are their phases. In this manner we can arrive to the following system:
\begin{subequations}\label{amplphaseeqs}
\begin{align}
&\dot{Q}_S=\gamma Q_S\sin(\Delta\epsilon),\label{amplphaseeqs1}\\
&\dot{Q}_N=\gamma Q_N\sin(\Delta\epsilon),\label{amplphaseeqs2}\\
&\dot{\epsilon}_S=-kc_S-\gamma\frac{Q_N}{Q_S}\cos(\Delta\epsilon),\label{amplphaseeqs3}\\
&\dot{\epsilon}_N=-kc_N+\gamma\frac{Q_S}{Q_N}\cos(\Delta\epsilon)\label{amplphaseeqs4},
\end{align}
\end{subequations}
where $\Delta\epsilon=\epsilon_N-\epsilon_S$ is the phase difference between the two CRWs. From eqs. \ref{amplphaseeqs} we can understand that (i) the waves can grow just with interacting between each other and (ii) the interaction strength depends on both $\gamma$ and $\Delta\epsilon$. When the phase difference is between $0<\Delta\epsilon<\pi$ we are in the growing phase-locking configuration, while if $-\pi<\Delta\epsilon<0$ it is a decaying configuration \cite{heifetz2005relating}. 

\begin{figure}
\begin{tabular}{cc}
\hskip-0.2cm\scalebox{0.29}{\includegraphics{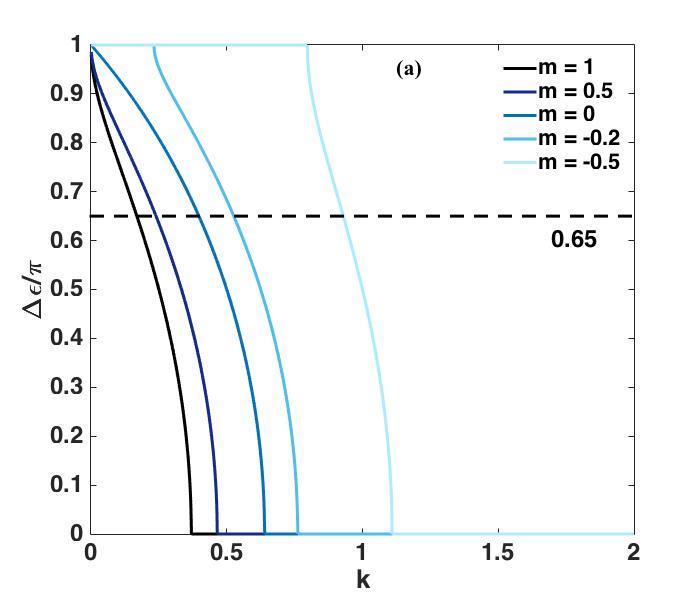}}&\hskip-0.5cm\scalebox{0.29}{\includegraphics{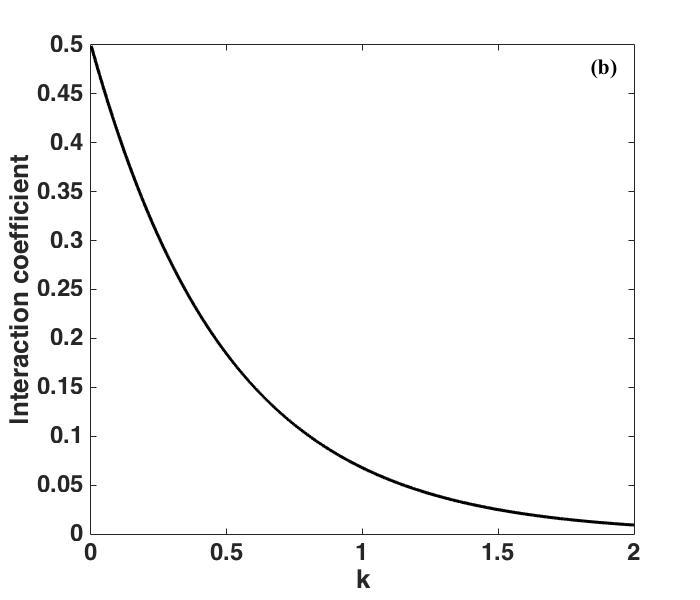}}\\
\end{tabular}
\caption{(a) Phase difference $\frac{\Delta\epsilon}{\pi}$ and (b) interaction coefficient for $L_d=\infty$ and different values of $m$. In (a) the dashed line represent the phase difference corresponding to the maximum growth rate $\Delta\epsilon=0.65$ for $m=0$.}\label{phasediffintcoeff} 
\end{figure}

We can obtain amplitudes and phases with computing the eigenvectors $V_{i}=[v_{i,S},v_{i,N}|$ of the matrix $M$ (eq. \ref{qpiccolokbar}), where the index $i$ depicts a different mode. In particular, the amplitude of S-CRW is obtained with $Q_{S} = |v_{i,S}|$, where $v_{i,S}$ is the $S$-component of eigenvector $V_i$, while their phase is $\epsilon_{S}= \arg(v_{i,S})$. Similarly, $Q_N$ and $\epsilon_N$ can be obtained. Note that the choice of the index $i$ ($i=1$ or $i=2$) is not important to determine the amplitude and the phase of the waves. 

 In figure \ref{phasediffintcoeff} we illustrate the phase difference $\frac{\Delta\epsilon}{\pi}$ and the interaction coefficient $\gamma$ for $L_d=\infty$ and different values of $m$. The phase difference corresponding to the maximum growth rate $\Delta\epsilon_{opt}=0.65\pi$ for $m=0$ is highlighted by a dashed line. 
Then for $m=0$ the gravest normal mode is not for $\Delta\epsilon=\frac{\pi}{2}$ where we have the maximal instantaneous growth rate as seen in eqs. \ref{amplphaseeqs}a,b.
To explain this we have to consider the two conditions to have phase-locking \cite{hayashi1987stable}: the waves must have the same (i) amplitude $Q_S=Q_N$ and (ii) phase speed $\dot\epsilon_1=\dot\epsilon_2$. The latter conditions is not respected for $\Delta\epsilon=\frac{\pi}{2}$ since the waves cannot affect each other's phase speed (see eqs. \ref{amplphaseeqs}c,d) and then the interaction can occur just if $c_S=c_N$ that it is not true. 
 In particular if $-\frac{\pi}{2}<\Delta\epsilon<\frac{\pi}{2}$ the two waves hinder their self-propagation against the mean flow, conversely if $\frac{\pi}{2}<\Delta\epsilon<\frac{3}{2}\pi$ they help the counterpropagation \cite{heifetz2005relating,biancofiore2017understanding}. Phase locking is favoured when interaction hinders the CRW counterpropagation.
The maximum modal growth rate occurs where there is a trade-off between the exponential increase of the interaction coefficient (see fig. \ref{phasediffintcoeff}b) at small wavenumbers and the need of a hindering configuration to enable phase-locking. The reader is referred to Ref. \onlinecite{heifetz2005relating} for more details on the CRW phase-locking.

The interaction coefficient does not depend on the presence of the background shear, while the phase difference is modified by $m$, see figure \ref{phasediffintcoeff}a. The optimal phase difference $0.65\pi$ is shifted towards long (short) wavelengths for positive (negative) $m$. This can explain why the instability is moved by a favorable/adverse shear towards long/short wavelengths as observed in fig. \ref{growthratem}b. Since the interaction coefficient does not change with introducing the background shear, the optimal phase difference corresponds now to a stronger/weaker interaction depending on the sign of $m$ as it is possible to see in fig. \ref{phasediffintcoeff}b. It should be noticed that the introduction of the background shear will modify the value of the optimal phase difference since the phase speed of the isolated CRWs depends on $m$ (see fig. \ref{growthratem}b). However we can assume that the dependence is not significant at least from a qualitative point of view. This point will be discussed later in more details.

\begin{figure}
\begin{tabular}{cc}
\hskip-0.2cm\scalebox{0.29}{\includegraphics{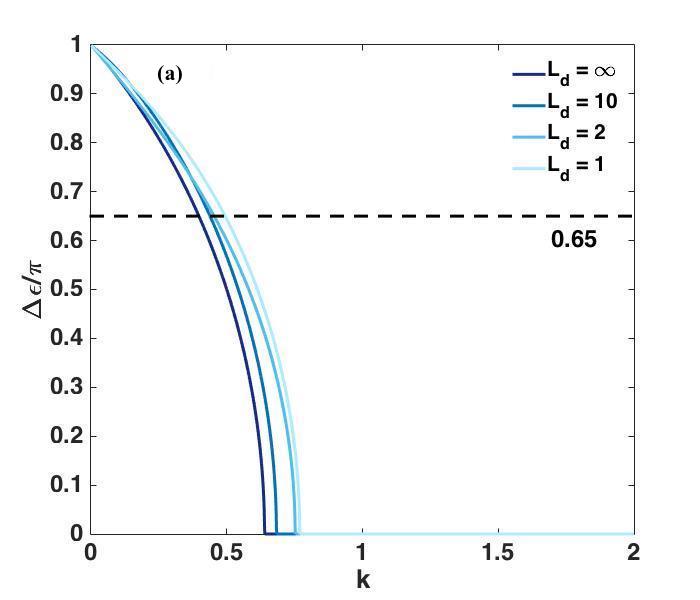}}&\hskip-0.5cm\scalebox{0.29}{\includegraphics{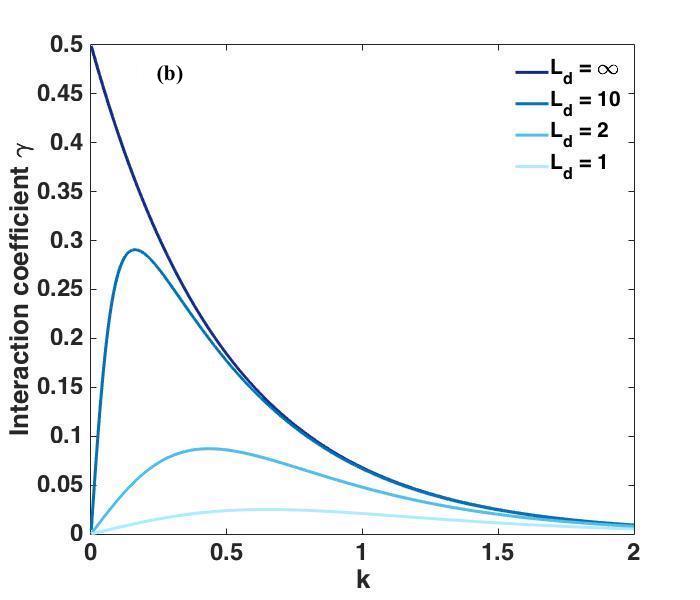}}
\end{tabular}
\caption{(a) Phase difference $\frac{\Delta\epsilon}{\pi}$ and (b) the interaction coefficient $\gamma$ vs wavenumber for different values of $L_d$ and $m=0$. 
The dashed line shows the value of the optimal phase difference $0.65\pi$ for the Rayleigh model (so $m=0$ and $L_d=\infty$) \cite{heifetz2005relating}.}\label{interactionm}
\end{figure}

In figure \ref{interactionm} we show (a) the phase difference $\frac{\Delta\epsilon}{\pi}$ and (b) the interaction coefficient $\gamma$ for $m=0$ and different values of the Rossby deformation radius. The phase difference is not strongly modified by introducing QG effects however the CRW interaction becomes weaker. This is in agreement with the behavior of the growth rate observed in figure \ref{growthrateLd}a: $L_d$ significantly damps the instability but the range of unstable wavenumber is almost untouched.

\begin{figure}
\begin{tabular}{cc}
\hskip-0.2cm\scalebox{0.29}{\includegraphics{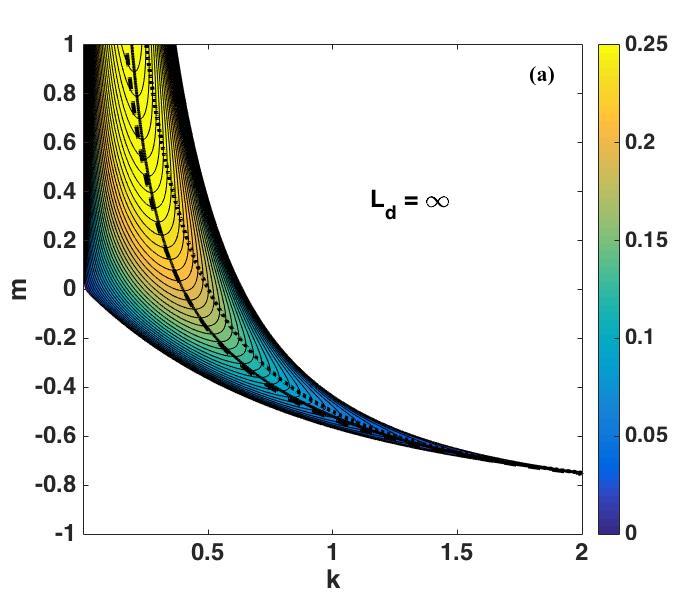}}&\hskip-0.5cm\scalebox{0.29}{\includegraphics{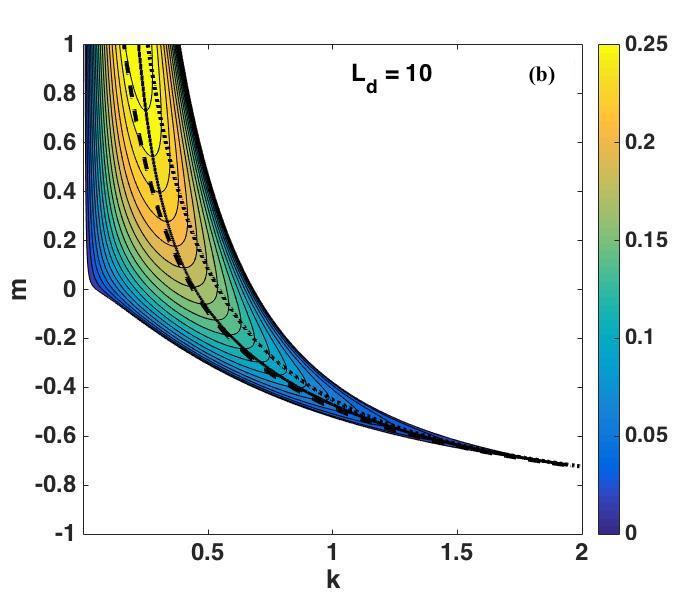}}\\
\hskip-0.2cm\scalebox{0.29}{\includegraphics{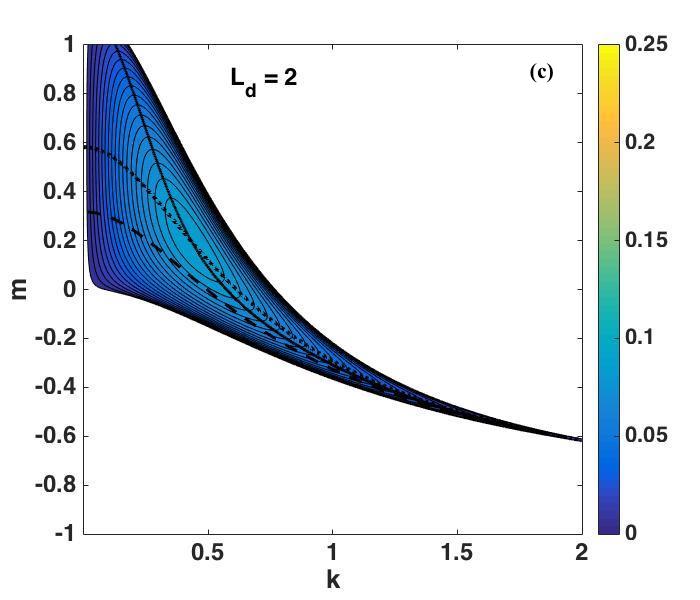}}&\hskip-0.5cm\scalebox{0.29}{\includegraphics{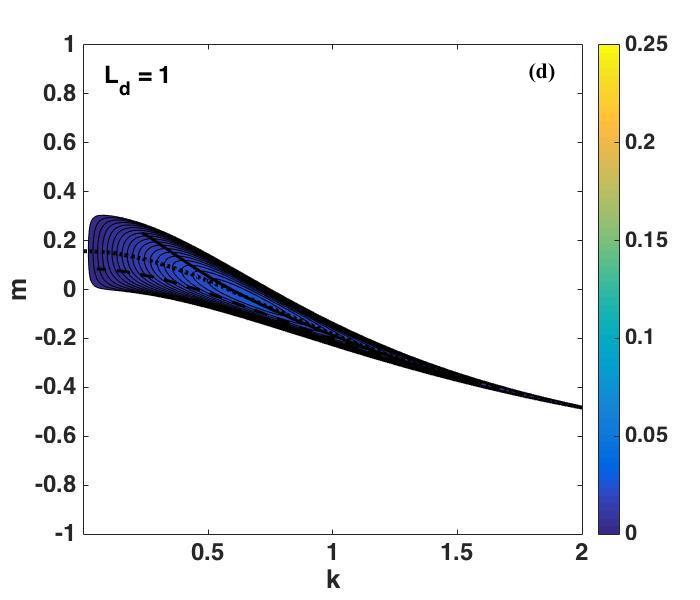}}
\end{tabular}
\caption{Growth rate $\omega_i$ in the plane $k$-$m$ for (a) $L_d=\infty$, (b) $L_d=10$, (c) $L_d=2$ and (d) $L_d=1$. The continuous line represents $k_{max}(m)$, the dashed line represents $k_{0.65\pi}(m)$ and the dotted line represents $k_{equal}(m)$.
}\label{plankmLdlines}
\end{figure}

In figure \ref{plankmLdlines} we show again the contours of the growth rate in the plane $k$-$m$ as in figure \ref{plankmLd} but this time comparing $k_{max}(m)$, i.e. the wavenumber corresponding to the maximum growth rate (depicted by continuous bold lines) with (i) $k_{0.65\pi}(m)$, i.e. the wavenumber for which the phase difference is equal to the optimal phase difference for $m=0$ and $L_d=\infty$: $0.65\pi$ (dashed lines) and (ii) $k_{equal}(m)$ (dotted lines). For all the values of the Rossby deformation radius, the unstable regions enclose the dotted lines, i.e. where the two isolated CRWs have the same phase speed. This confirms that the two isolated CRWs must have a similar phase speed to have phase-locking since the change in phase speeds due to the presence of the other CRW is small. For $L_d=\infty$ (a) we observe that the position of the maximum of the growth rate almost fully corresponds to where the phase difference is equal to $\Delta\epsilon=0.65\pi$. This shows that the introduction of the background shear does not qualitatively modify the phase-locking. However the discrepancy between the dashed and the continuous lines increases as much as $L_d$ is reduced. Therefore QG effects significantly modify the value of the optimal phase difference. 

\begin{figure}
\begin{center}
\scalebox{0.31}{\includegraphics{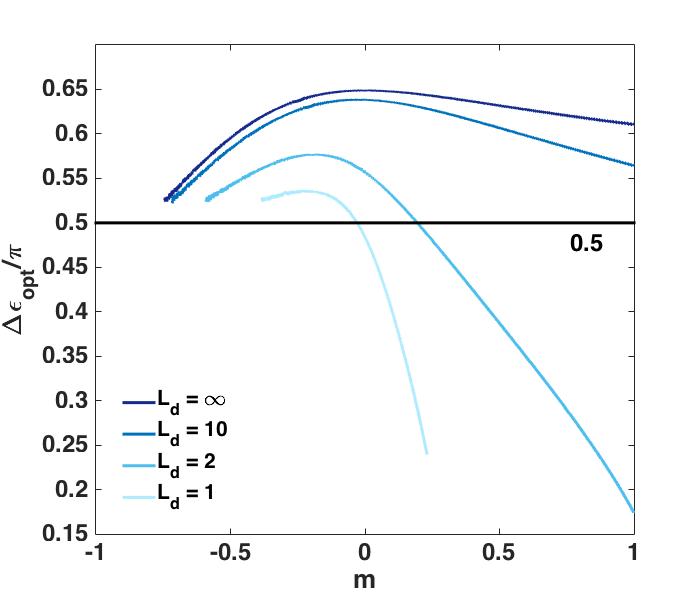}}
\caption{Phase difference corresponding to the maximum growth rate $\frac{\Delta\epsilon_{max}}{\pi}$ vs wavenumber for $m=0$ and different values of $L_d$. The black line represents the limit between helping and hindering configuration.
}\label{diffepsmax}
\end{center}
\end{figure}

To make this point more evident we illustrate in figure \ref{diffepsmax} the value of the phase difference corresponding to the maximum growth rate $\Delta\epsilon_{opt}$. The black line discriminates between helping ($-\frac{\pi}{2}<\Delta\epsilon<\frac{\pi}{2}$) and hindering ($\frac{\pi}{2}<\Delta\epsilon<\frac{3}{2}\pi$) configurations. When the Rossby deformation radius is infinite the influence of the background shear is weak. However the optimal phase difference is slightly diminished by adding both a favorable or an adverse shear. This means that either the influence of one CRW on the other's phase speed is stronger or the two isolated CRWs have a closer value of the phase speed so they need less hindering effect to phase-lock. The former case occurs for $m>0$ since also the influence on the each other's phase speed is proportional to $\gamma$ which is larger at small wavenumbers (see figure \ref{phasediffintcoeff}b). For $m<0$ instead the latter case occurs since the difference between the two phase speeds of the isolated CRWs is smaller. 

However the dependence of $\Delta\epsilon_{opt}$ on $L_d$ is more significant, in particular for positive $m$. We observe a general decrease in the optimal phase difference. This is due to the fact that the interaction is weaker when QG effects are present as seen in fig. \ref{interactionm}b. For a favorable background shear the two CRWs need a helping configuration to phase-lock if QG effects are strong. This occurs since the two CRWs lose their counter-propagative character for small $L_d$, as observed in fig. \ref{growthrateLd}b. Thus, to have phase-locking, each CRW needs help from the other wave to withstand the mean flow.

\section{Non-Linear Simulations}\label{NLS}

In this section we conduct non-linear simulations of the quasi-Rayleigh profile which have presented in section \ref{justmodel}. 
The simulations will be setup to see what happens when a shear layer, treated as a filament with unit PV in its interior and zero PV on its exterior, is subject to an additional global background forced shear.   Strictly speaking, the model setup which we examine here is slightly different than the theoretical one considered in section 2, where the background shear is not externally forced (per se).    
In order to establish some quantitative correspondence between these two setups some care must be applied in constructing the initial profile to be numerically evolved.  In particular, it will be of value to verify that the linear evolution phase predicted from the idealized model of section 2 is realized in some meaningful way through the initial setup and subsequent evolution of the nonlinear numerical solutions described here.   To this end we assert the following equivalence conjecture:
\par
\medskip
\noindent{\bf Conjecture
(An equivalence conjecture)}
{\it Within a given QG framework where $L_d$ is the same, the linear evolution of two constant PV filaments of the same width will be the same provided the two filaments exhibit both (i) the same jump in PV across their boundaries and,
(ii) the same mean streamwise velocities on their boundaries.}
\par
\medskip

This proposition is an immediate implication of the counterpropagating Rossby wave perspective.  Indeed, the linear analysis of section 2 demonstrates that the scale and growth rate of a shear layer of given filamentary width and with constant interior PV depends only on three things: the value of $L_d$, the value of the mean velocities on the two boundaries of the shear filament layer, and the jump in PV across filament boundaries.

As such, this section is subdivided by: section \ref{numericalsetup} which describes the numerical method we use, 
revisits the governing equations and details the initial profiles we setup for our numerical experiments;
in section \ref{validation} we illustrate the tests we perform to validate our numerical code and demonstrate correspondence to
the linear theory of section 2; while in section \ref{results} the results of our numerical experiments are presented and discussed.

\subsection{Governing equations and numerical setup}\label{numericalsetup}

\subsubsection{Governing equations}
The nonlinear equations of motion we solve are
\beq 
\left[ \frac{\partial}{\partial t} + (\tilde my+u)\frac{\partial}{\partial x}
+ v\frac{\partial}{\partial y}\right] q = -\nu\sub 8 \nabla^8 q,
\label{num_eqns}
\eeq 
where
\beq 
q \equiv \nabla^2 \psi - L_d^{-2}\psi; \qquad
u \equiv \frac{\partial \psi}{\partial y}, \quad
v \equiv -\frac{\partial \psi}{\partial x}.
\label{num_eqns_supplemental}
\eeq 
We have set the planetary rotation parameter $\beta$ to zero.  The background shear ($\tilde my$) is given as immutable and is the only source of continual external forcing. This is in addition to the original filament profile with which we initiate the simulation.
Deviations of this flow are given
by the initial form adopted for $q$.  We will consider ``rounded" models of the basic forms described in Eq. (\ref{basicvorticity1}), see further below.  Because regular viscosity is set to zero, in order to dissipate enstrophy/energy at the smallest scales of the simulation we apply an 8th order superviscosity operator $-\nu_8 \nabla^8$.    The connection between $\tilde m$ and the value of $m$ used in our analysis in section 2 will be clarified further in subsection \ref{simulationsetup}.

\subsubsection{Numerical method}
The set of equations (\ref{num_eqns}-\ref{num_eqns_supplemental}) will be solved using standard pseudospectral methods on a doubly periodic domain.  An example of the output of the code used here can be found in the chapter on turbulence describing 2D decaying processes in Regev {\it et al} \cite{regev2016modern}.
This means to say that at a given time step $t_n$ the PV is represented as
\beq
q(x,y,t_n)  = \sum_{\ell=0}^{N_y}\sum_{k=-N_x}^{N_x} q\sub{\ell,k}^{n} \exp\left[{\frac{2\pi i k x}{L_x} + \frac{2\pi i\ell y}{L_y}}\right]+ {\rm c.c.}
\eeq
in which $\ell,k$ and $N_x,N_y$ are integers.   The physical size of the domain in the x and y directions, respectively, are $L_x,L_y$.  
$q\sub{\ell,k}^{(n)}$ is the amplitude of the Fourier component at time step $n$, as indicated by a superscript.
All derivative operations are assessed in Fourier space while the nonlinear advection terms are assessed in physical space using a standard 2/3 dealiasing rule. 
The streamfunction is the solution of the corresponding Poisson equation found in (\ref{num_eqns_supplemental}).  In spectral space this becomes a simple algebraic
relationship.  Thus, the Fourier component of the streamfunction is given by
\beq
\psi\sub{\ell,k}^{n} = -\frac{q\sub{\ell,k}^{(n)}}{K^2\sub{k,\ell} + {L_d^{-2}}},
\eeq
where the total wavenumber $K\sub{k,\ell}$ is defined as
\beq
K^2\sub{k,\ell} \equiv \frac{4\pi^2 k^2 }{L_x^2} + \frac{4\pi^2 \ell^2 }{L_y^2}.
\eeq
The velocity fields, and all other derivatives, are assessed in spectral space via simple multiplication, e.g.,
\beq
u\sub{\ell,k}^{n} = -  \frac{2\pi i\ell }{L_y}\cdot \psi\sub{\ell,k}^{(n)} \qquad {\rm and} \qquad
v\sub{\ell,k}^{n} =   \frac{2\pi i k}{L_x}\cdot \psi\sub{\ell,k}^{(n)}.
\eeq
The PV is evolved in Fourier space.  The hyper-viscosity is applied during the time step scheme based on the modified Crank-Nicholson method described in both Refs. \onlinecite{umurhan2004hydrodynamic} and \onlinecite{tamarin2015nonnormal}.  
According to this procedure time stepping routine then predicts the updated PV amplitude at time step $n+1$ via,
\beq
q\sub{\ell,k}^{n+1} = e^{-2 \delta t \nu\sub 8 |K\sub{k,\ell}|^8} q\sub{\ell,k}^{n-1} + 2 \delta t e^{-\delta t \nu\sub 8 |K\sub{k,\ell}|^8} {\rm N}\sub{\ell,k}^{n},
\eeq
where ${\rm N}\sub{\ell,k}^{n}$ is the aforementioned nonlinear advection term in Fourier space.   Most simulations were run $N_x = 1120$ and $N_y =  1280.$  With 2/3 dealiasing rule this means there were $2\times800\times800 \approx 1.3\times 10^6$ active Fourier modes being evolved. In typical runs we adopted values of $L_x =49.2928$ and $L_y = 60$ which amounts to a typical grid spacing of about $\Delta x \approx 0.04$ and $\Delta y \approx 0.05$.  For values of $L_d$ that were less than 10 we considered $L_y = 30$ instead. This adjustment was chosen since for small values of $L_d$ means that individual vortex structures have much more limited physical ``reach".  Also this choice was motivated by the practical observation wherein we find that runs involving small values of $L_d$ rarely spread in the y-direction. In these smaller $L_d$ runs we continued  to keep the same number of Fourier modes in the y-direction. 
 The choice of $\delta t$ typically was either $0.005$ or $0.0025$.  In some rare instances, when $|m| = 0.5$, we found that we had to reduce the step size to as little as $\delta t = 0.001$ in order to avoid violating CFL constraints.  
\par 
We usually choose values of $\nu_8$  in the vicinity of $ 1/(256 \times 10^{11})$.
This means that for the highest resolved Fourier mode (with, say, $\delta t = 0.0025$ with $k=800,\ell = 800$) the hyperviscosity reduces the amplitude by shaving off about $0.65\%$ of the amplitude of the mode since
$\exp\left[{- \delta t \nu\sub 8 |K\sub{k,\ell}|^8}\right] = 0.9935$.  By consequence, via the process of hyper-viscosity alone, the highest resolved wavemode has an e-folding timescale roughly of about $\Delta t = 0.5$.  Raising the value of $\nu\sub 8$ had visibly noticeable effects: higher values  enhanced an effective stickiness between well formed vortices and tended to cause mergers to happen more readily than otherwise.  
  Too small a value of $\nu_8$ usually results in the pile-up of enstrophy and energy at the cutoff wavenumbers (corresponding to that selected by 2/3 dealiasing rule) -- in those cases the numerical experiments blew-up via the generation of $2\Delta x$ waves.  We therefore settled upon the value of $\nu_8$ where we found that the observed inertial spectrum showed little to no change and the numerical experiments were free of $2\Delta x$ instabilities.

\subsubsection{Simulation setup}\label{simulationsetup}
We must setup the initial filament profile as a streamwise uniform solution of the fundamental equations.  As per our considerations of the beginning of this section, in order to connect to the filaments examined in the linear analysis of section 2 we have to make sure that the jump in PV between across the filament boundaries are the same -- in this case this jump in PV is $\Delta q = 1$.  Furthermore, the streamwise velocities on the two boundaries must be the same as well.   Constructing this consistently will determine a connection formula between $\tilde m$ and $m$. 
We approach in the following way.  We divide the PV as one being composed of a piecewise constant ``quasi-Rayleigh" profile plus the aforementioned globally forced background uniform shear, $\tilde m y$.
This quasi-Rayleigh PV is denoted by $Q\sub {qR}$ and is the solution of 
\beq
\left(\nabla^2 - \frac{1}{L_d^2}\right)\Psi\sub {qR} \equiv Q\sub{qR} =   \left\{ \begin{array}{rll}0,&{\rm for} &
|y|>1;\\ -1, &{\rm for}&
|y|\le 1,
\end{array}\right.
\label{qR_def}
\eeq
where $\Psi\sub{qR}$ is the corresponding streamfunction of this steady state.
Comparison of the right-hand-side of eq. (\ref{qR_def}) to the PV adopted  in section 2 -- i.e., as summarized in eq. (\ref{basicvorticity1}) -- shows that the two
show the same jump in PV across their filament boundaries at $y=\pm 1$.
The mean streamwise velocity of this quasi-Rayleigh profile is given as
\beq
\displaystyle
U\sub{qR}(y)=\left\{ \begin{array}{lll}
u_0 e^{\frac{1-y}{L_d}}
&{\rm for} &
y>1\\ 
u_0\sinh\left(\frac{y}{L_d}\right)/\sinh\left(\frac{1}{L_d}\right) &{\rm for}&
|y|\le 1 \\
-u_0 e^{\frac{y+1}{L_d}}
&{\rm for} &
y<-1
\end{array}\right.
\qquad
{\rm with} \ \ 
u_0 = \frac{L_d}{1+\coth\left[\frac{1}{L_d}\right]};
\label{basicvelocity3}
\eeq
To complete the correspondence between the flow setup here and that considered in section 2, we must make sure that the 
velocities at $y=\pm 1$ are equal to one another.  This means setting $\pm \tilde m +U\sub{qR}(\pm 1) = U(\pm 1)$
where $U$ is the streamwise velocity found in eq. (\ref{basicvelocity1}).  Owing to the symmetry of the profile with respect to $y=0$, this amounts to
\beq
\displaystyle
\tilde m+ \frac{L_d}{1+\coth\left[\frac{1}{L_d}\right]}
= L_d(m+1) e^{-\frac{1}{L_d}}\sinh\left[\frac{1}{L_d}\right],
\eeq
simplifying this relationship for $\tilde m = \tilde m(m,L_d)$, we find
\beq
\tilde m(m,L_d)  = \frac{m L_d }{1 + \coth\left(1/L_d\right)},
\eeq
which completes the correspondence we have sought to establish.\par
\bigskip
All simulations are then initially seeded with these piecewise constant quasi-Rayleigh filament solutions, i.e., $q(t=0) = Q\sub{qR}$.  
We do this by utilizing a ``rounded" 
approximation of $ Q\sub{qR}$ given by $\tilde  Q\sub{qR}$ where
\beq
\tilde Q\sub{qR}(\epsilon) = -\frac{1}{2}\left(\tanh\left[\frac{y+1}{\varepsilon}\right] - 
\tanh\left[\frac{y-1}{\varepsilon}\right]\right),
\eeq  
in which $\varepsilon$ controls the tightness of the transition across the filament boundaries.  
A cursory inspection shows that $$\lim_{\varepsilon \rightarrow 0}\tilde Q\sub{qR}(\epsilon)
\rightarrow Q\sub{qR}$$.
In order to 
resolve the PV transition at $y=\pm 1$ we use at least 5-7 grid points, which generally meant having values of $\varepsilon = 0.05$ (keeping in mind a typical grid spacing of about $\Delta y \approx 0.05$).  Atop this
initial filament we introduce an additional amount of white noise in the PV field.  This noise is constrained to be non-zero in a region $|y|< 2$.  At maximum, the noisy PV field has an amplitude which is about 25$\%$ of the value inside the filament.

\subsection{Validation}\label{validation}
\begin{figure}
\begin{tabular}{cc}
\hskip-0.1cm\scalebox{0.141}{\includegraphics{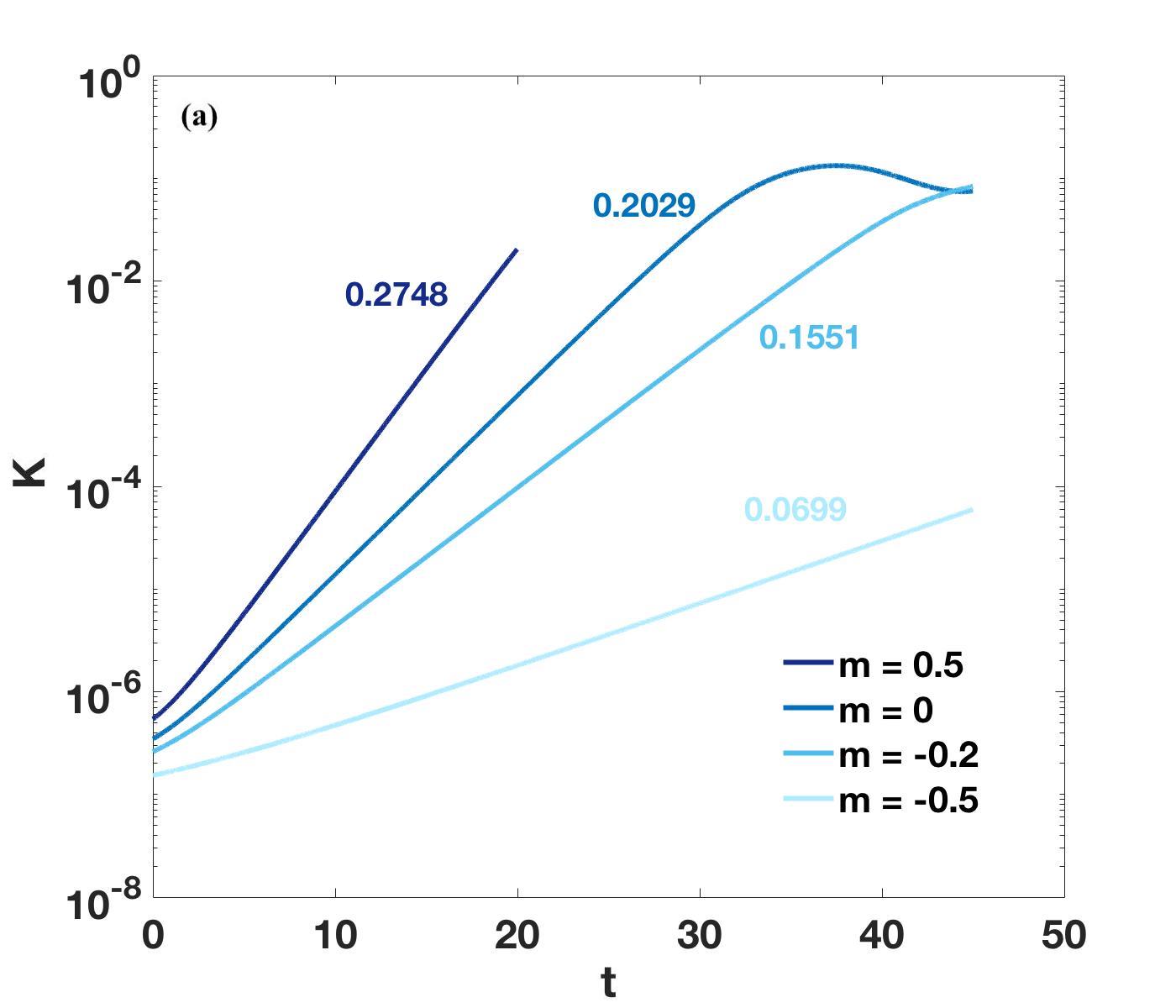}}&\hskip-0.2cm\scalebox{0.141}{\includegraphics{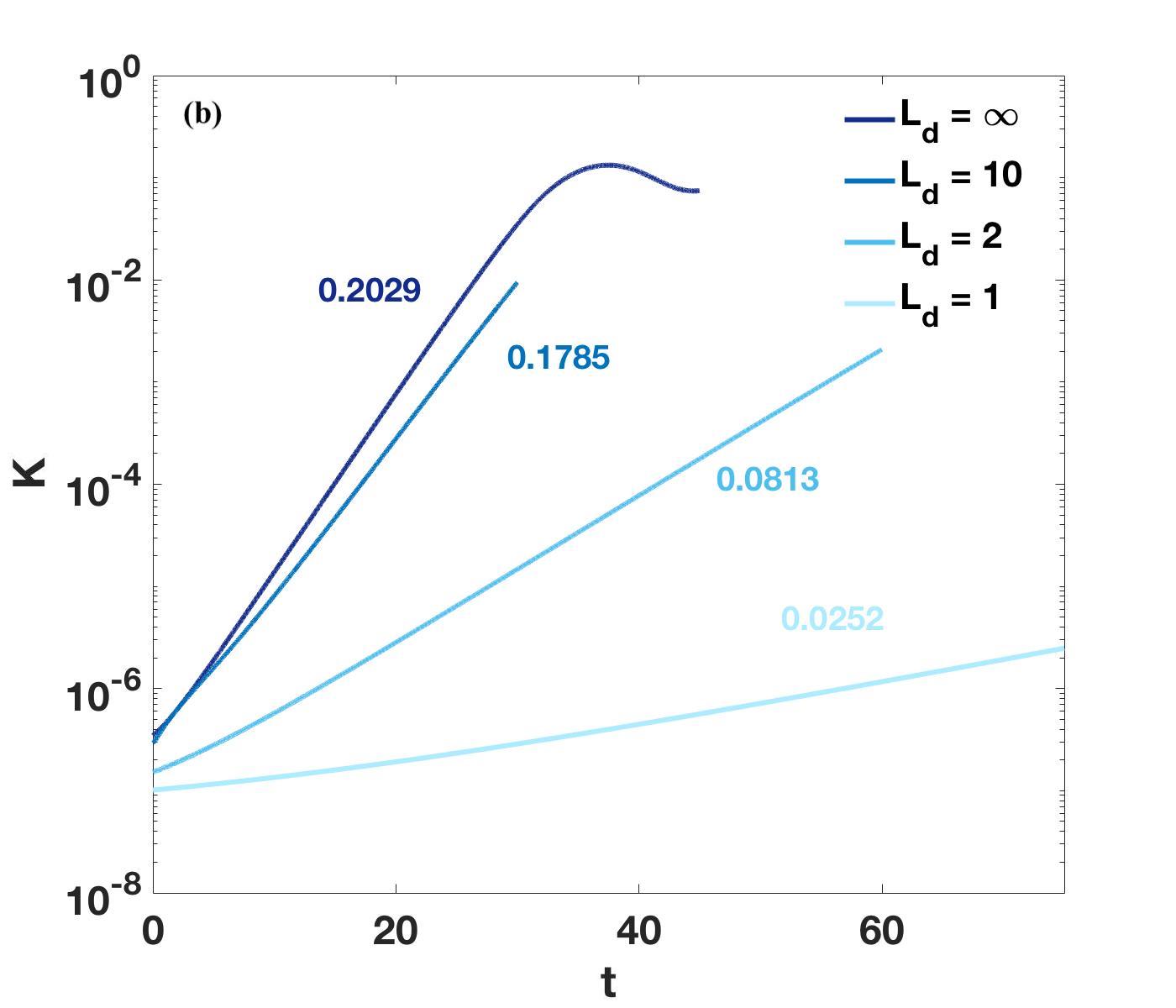}}
\end{tabular}
\caption{Temporal behavior of the spatially averaged perturbation kinetic energy $K$ for (a) $L_d=\infty$ and different values of $m$ and (b) $m=0$ and different values of $L_d$. Note that the domain sizes are $L_x=\frac{2\pi}{k_{max}}$ (see table \ref{tab10} for the values of $k_{max}$) and the domain height is $L_y=60$ for all the configurations but for $L_d\leq2$ in which $L_y=30$.}\label{DNSgrowthrate}
\end{figure}

We validate the behavior of the simulations by assessing the growing phase in the experiments against the linear theory predictions of section \ref{LSA} while keeping in mind the equivalence conjecture posed at the beginning of section {\ref{NLS}}. For given values of $m$ we initiate simulations with the PV fields described above by correctly setting up the PV field with the numerically equivalent value of the background forcing $\tilde m$. To this PV field we add an additional small amplitude ($\sim0.001$) perturbation corresponding to the fastest growing mode determined from linear theory. This entails sinusoidally perturbing the two boundaries of the filament while making sure that the wavelength of the maximally growing mode, $k_{max}$ and the difference in the phase of the two waves corresponded to those predicted for the maximally growing mode, i.e., $\Delta \epsilon = \Delta \epsilon_{max}$, where $\Delta \epsilon_{max} = \Delta \epsilon_{max}(m,L_d)$ is determined from the analysis in section \ref{LSA}.  
The streamwise domain size of the numerical experiment was chosen to fit exactly one 
maximally growing mode, i.e., $L_x = 2\pi/k_{{\rm max}}$.
The linear theory predicts a growth rate for this mode given by $\omega_i\left({\rm LIN}\right)$.  We turn off hyper-viscosity in its entirety and run the simulation until the profile transitions into the nonlinear regime, where upon we end the numerical experiment. In the temporal window in which the disturbances are clearly in the linear range, we extract the perturbation kinetic energy field from the solutions and calculate an average perturbation kinetic energy $K$ -- see Figure \ref{DNSgrowthrate}.  During the linear growth phase this kinetic energy grows by 4-5 orders of magnitude before saturating.  From this time series we do a linear regression fit and extract a growth rate
which we call $\omega_i\left({\rm DNS}\right)$.
In Table \ref{tab10} we summarize the results of these validations. We find that the predicted and measured values of $\omega_i$ differed from one another by less than 1\% for all values except two simulation runs: (i) $m=0.5$ and $L_d = 2$ and (ii) $m=0$ and $L_d=1$ which showed errors around 1.2\%.
Given the good correspondence between the predicted growth rates and that observed in the numerical experiments we are confident that the numerical simulations are good and, further, that the equivalence conjecture we have posed is a robust concept.

\begin{table}
\vskip 0.5cm
\center\begin{tabular}{cccccccc}
\hline
$m$ & $L_d$ &$\tilde m$ &$k_{max}$ & $\Delta\epsilon_{max}\cdot\pi^{-1}$ &$\omega_i$ (LIN) & $\omega_i$ (DNS) & Err \% \\
\hline
$0.5$    & $\infty$ & $0.500$ & $0.252$  & $0.6321$ &$0.2763$ & $0.2748$ & $0.53$ \\
\hline
$0$   & $\infty$ & $0$ & $0.398$  &$0.6492$& $0.2012$ & $0.2029$ & $0.8$ \\
\hline
$-0.2$  & $\infty$ & $-0.200$ & $0.532$  &$0.6419$& $0.1557$ & $0.1551$ & $0.38$ \\
\hline
$-0.5$  & $\infty$ & $-0.500$ & $0.956$  & $0.5961$ &$0.0705$ & $0.0699$ & $0.85$ \\
\hline
$0.5$    & $10$ & $0.453$ & $0.285$  & $0.6060$ &$0.2437$ & $0.2448$ & $0.45$ \\
\hline
$0$    & $10$ & $0$ & $0.448$  & $0.6382$ &$0.1768$ & $0.1785$  & $0.95$\\
\hline
$-0.5$    & $10$ & $-0.453$ & $1.065$  & $0.5834$ &$0.0566$ & $0.0571$  & $0.88$  \\
\hline
$0.5$    & $2$ & $0.316$ & $0.302$  & $0.3874$ &$0.0754$ &$0.0763$  & $1.19$\\
\hline
$0$    & $2$ & $0$ & $0.573$  & $0.5565$ &$0.0810$ & $0.0813$ & $0.37$ \\
\hline
$-0.5$    & $2$ & $-0.316$ & $1.491$  & $0.5439$ &$0.0202$ & 0.0202  & $0$ \\
\hline
$0$    & $1$ & $0$ & $0.592$  & $0.4843$ &$0.0249$ & $0.0252$ & $1.20$\\
\hline
\end{tabular}
\caption{Table summarizes the test cases run for validating the non-linear simulations and demonstrating the robustness of the equivalence conjecture.}\label{tab10}
\end{table}

\subsection{Results}\label{results}

In this section the results of the non-linear simulations are reported. We  analyze (i) the potential vorticity fields (section \ref{PVfields}), (ii) the momentum thickness growth (section \ref{momthicknesssection}) and (iii) the perturbation kinetic energy (section \ref{pertkinenergysection}). 
In our simulations the domain size is settled to $L_x=49.2928$ and $L_y=60$ ($L_y=30$ for $L_d\leq 2$ for ensuring numerical stability). A random disturbance is added to the initial field as explained in section \ref{simulationsetup}.

\subsubsection{Potential vorticity fields}\label{PVfields}

\begin{figure}
\begin{tabular}{cc}
\hskip-0.2cm\scalebox{0.095}{\includegraphics[trim={10.5cm 0cm 9.5cm 0cm},clip]{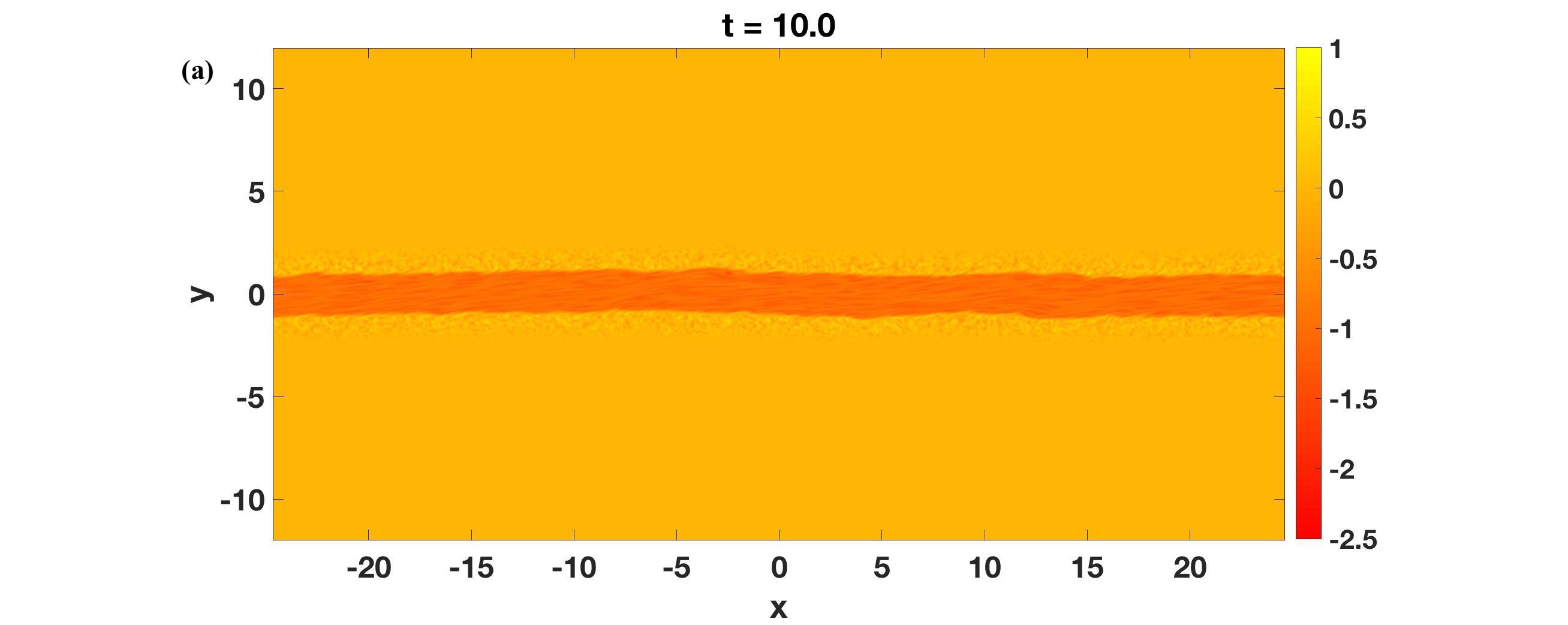}}&\hskip-0.2cm\scalebox{0.095}{\includegraphics[trim={10.5cm 0cm 9.5cm 0cm},clip]{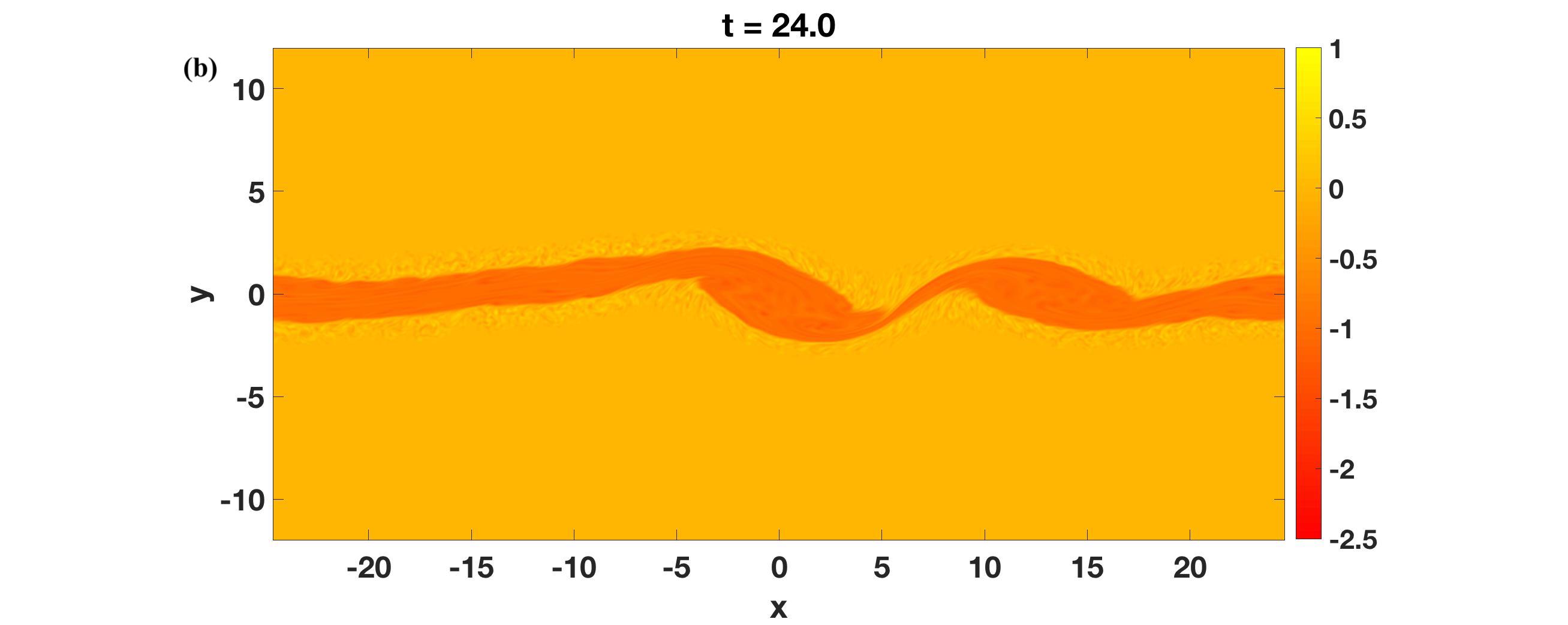}}\\
\hskip-0.2cm\scalebox{0.095}{\includegraphics[trim={10.5cm 0cm 9.5cm 0cm},clip]{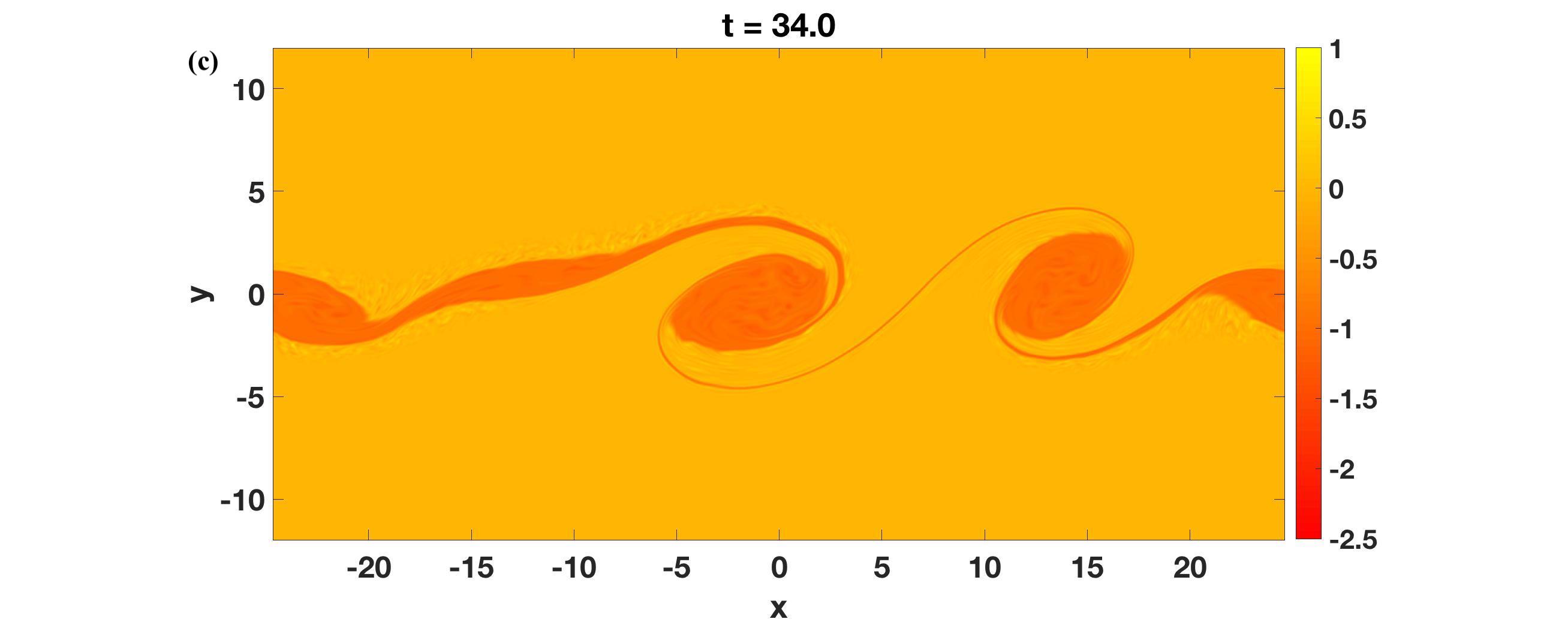}}&\hskip-0.2cm\scalebox{0.095}{\includegraphics[trim={10.5cm 0cm 9.5cm 0cm},clip]{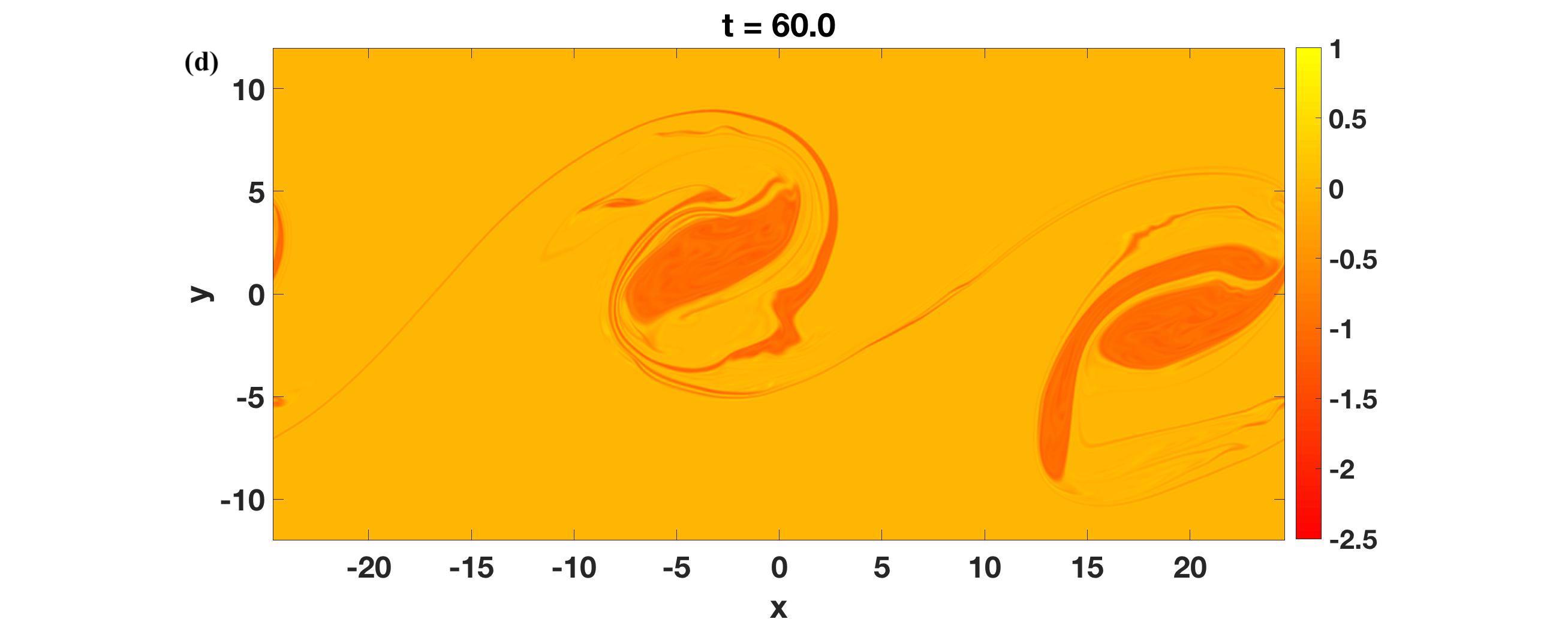}}\\
\hskip-0.2cm\scalebox{0.095}{\includegraphics[trim={10.5cm 0cm 9.5cm 0cm},clip]{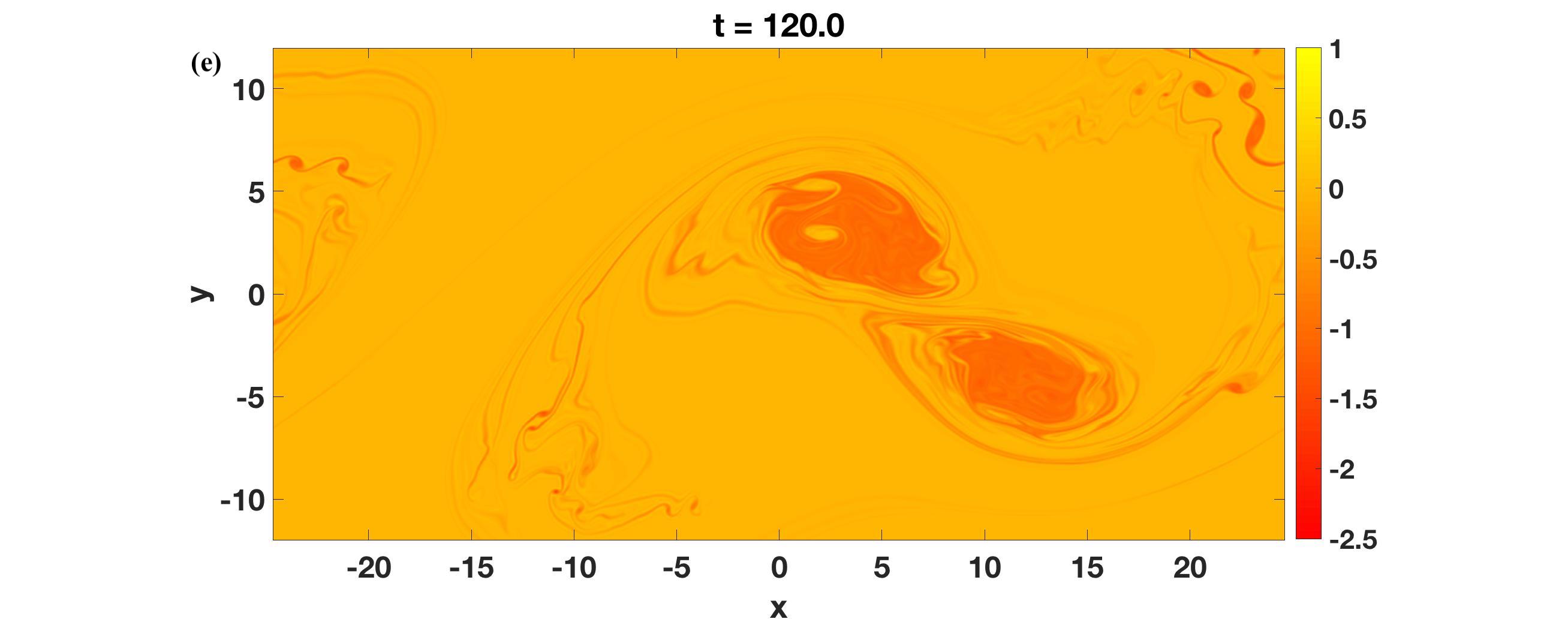}}&\hskip-0.2cm\scalebox{0.095}{\includegraphics[trim={10.5cm 0cm 9.5cm 0cm},clip]{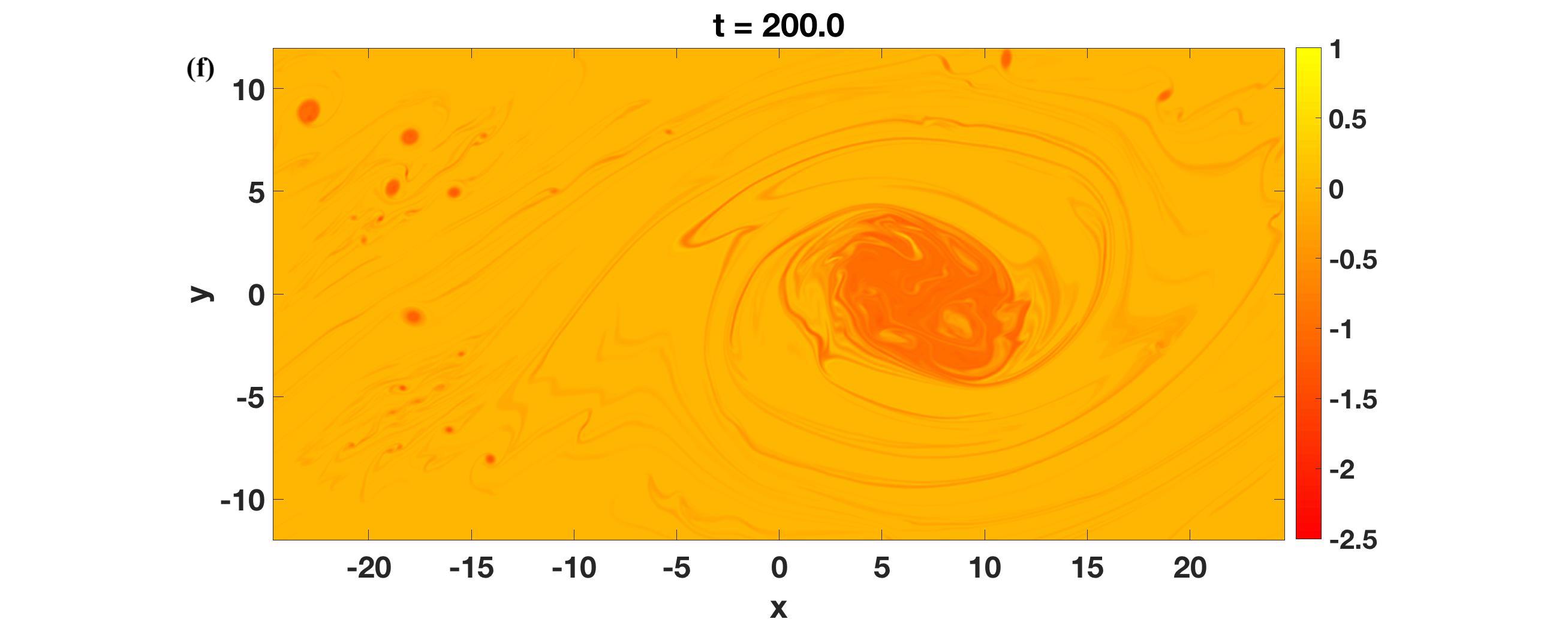}}
\end{tabular}
\caption{Potential vorticity fields for $m=0$ and $L_d=\infty$ at different times (a) $t=10$, (b) $t=24$, (c) $t=34$, (d) $t=60$, (e) $t=120$ and (f) $t=200$.}\label{snapshotsm0LdInfty}
\end{figure}

In figure \ref{snapshotsm0LdInfty} the potential vorticity fields showing the non-linear evolution of the classical Rayleigh profile (i.e. $m=0$ and $L_d=\infty$) are illustrated at different times from (a) $t=8$ to (f) $t=150$. Four vortices are created by the initial disturbance (b,c). These vortices pair between each other causing the growth of the shear layer (d,e). Finally they combine to form one single rotating vortex with filamentary arms (f).
The vortex pairing is the main cause of the spreading of the mixing layer \cite{winant1974vortex}.
This behavior is typical of two-dimensional mixing layers and the reader is referred to the review of Ho \& Huerre \cite{ho1984perturbed}. 

\begin{figure}
\begin{tabular}{cc}
\hskip-0.2cm\scalebox{0.095}{\includegraphics[trim={10.5cm 0cm 9.5cm 0cm},clip]{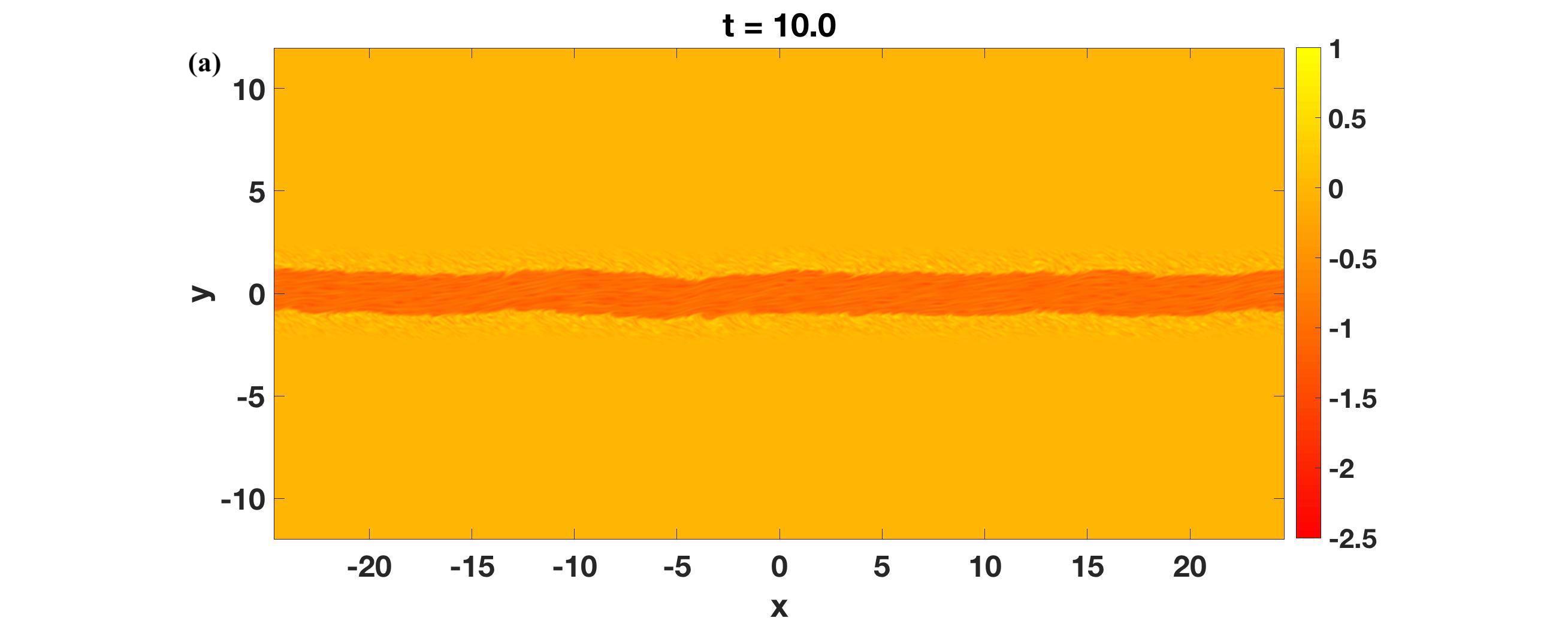}}&\hskip-0.2cm\scalebox{0.095}{\includegraphics[trim={10.5cm 0cm 9.5cm 0cm},clip]{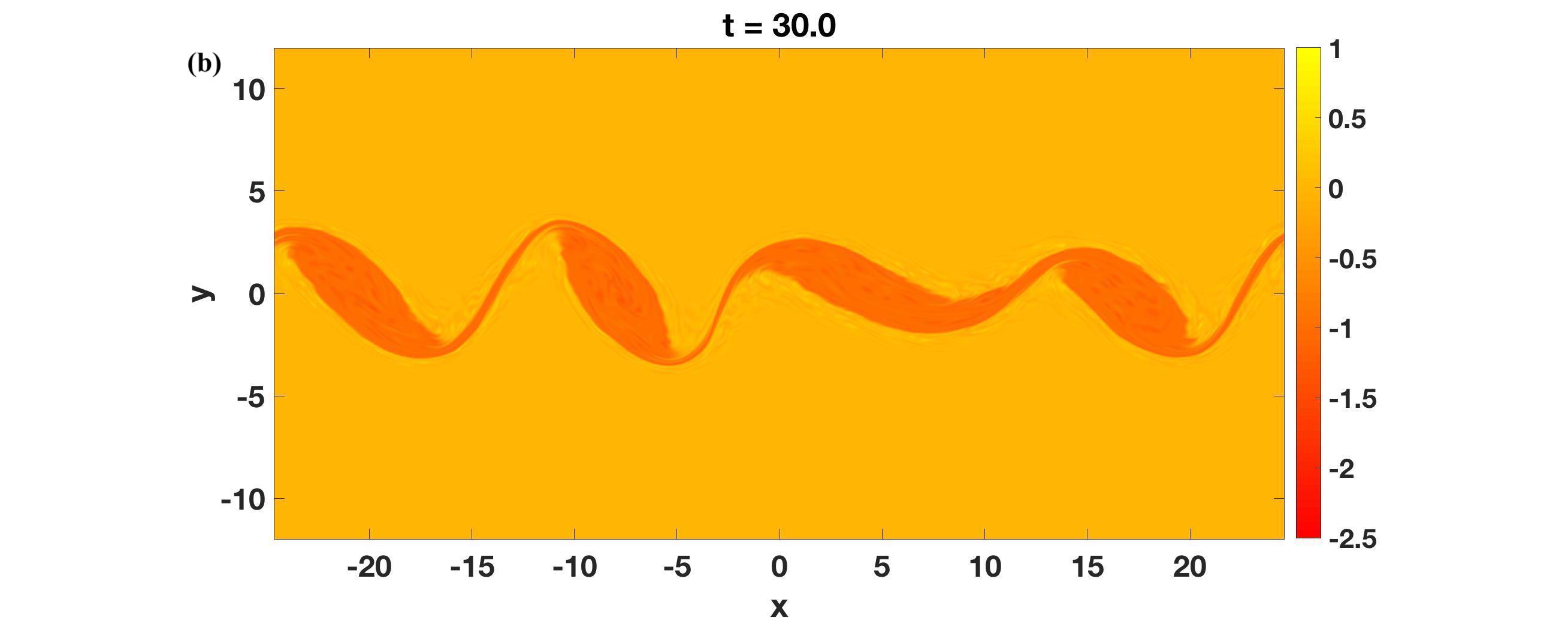}}\\
\hskip-0.2cm\scalebox{0.095}{\includegraphics[trim={10.5cm 0cm 9.5cm 0cm},clip]{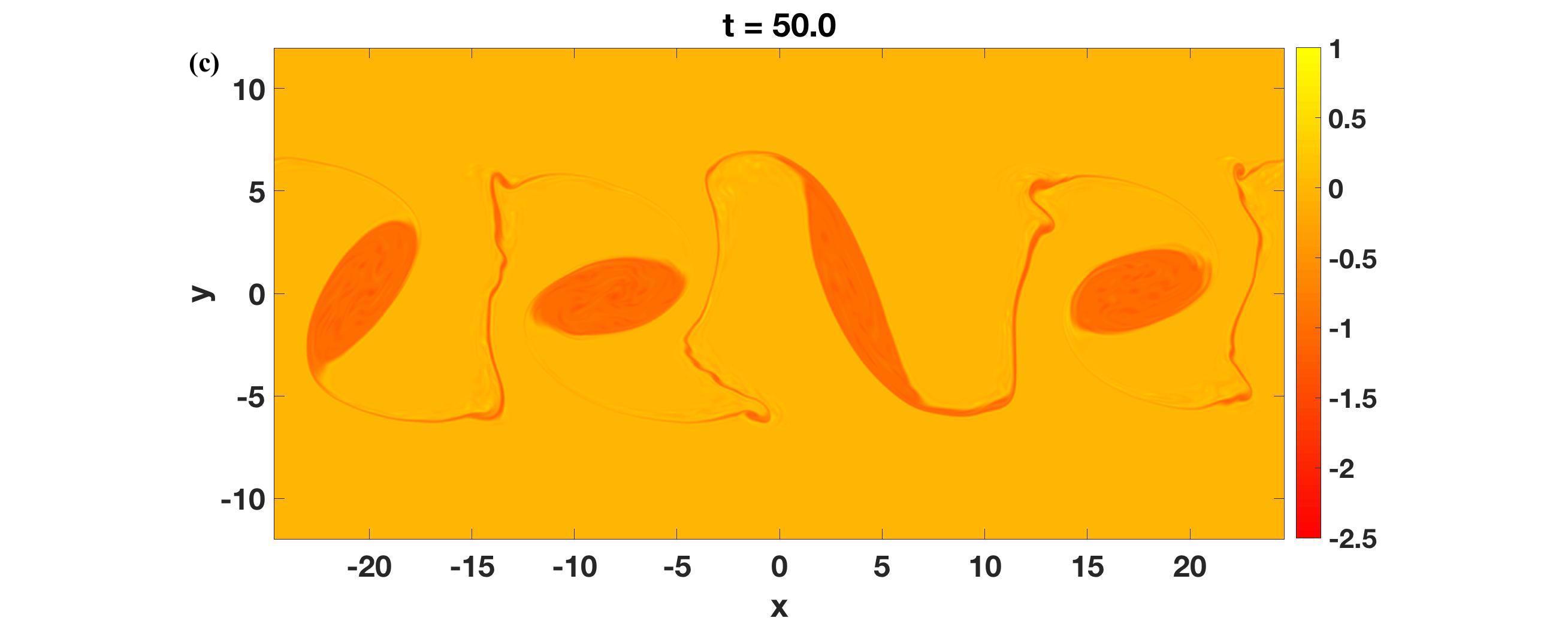}}&\hskip-0.2cm\scalebox{0.095}{\includegraphics[trim={10.5cm 0cm 9.5cm 0cm},clip]{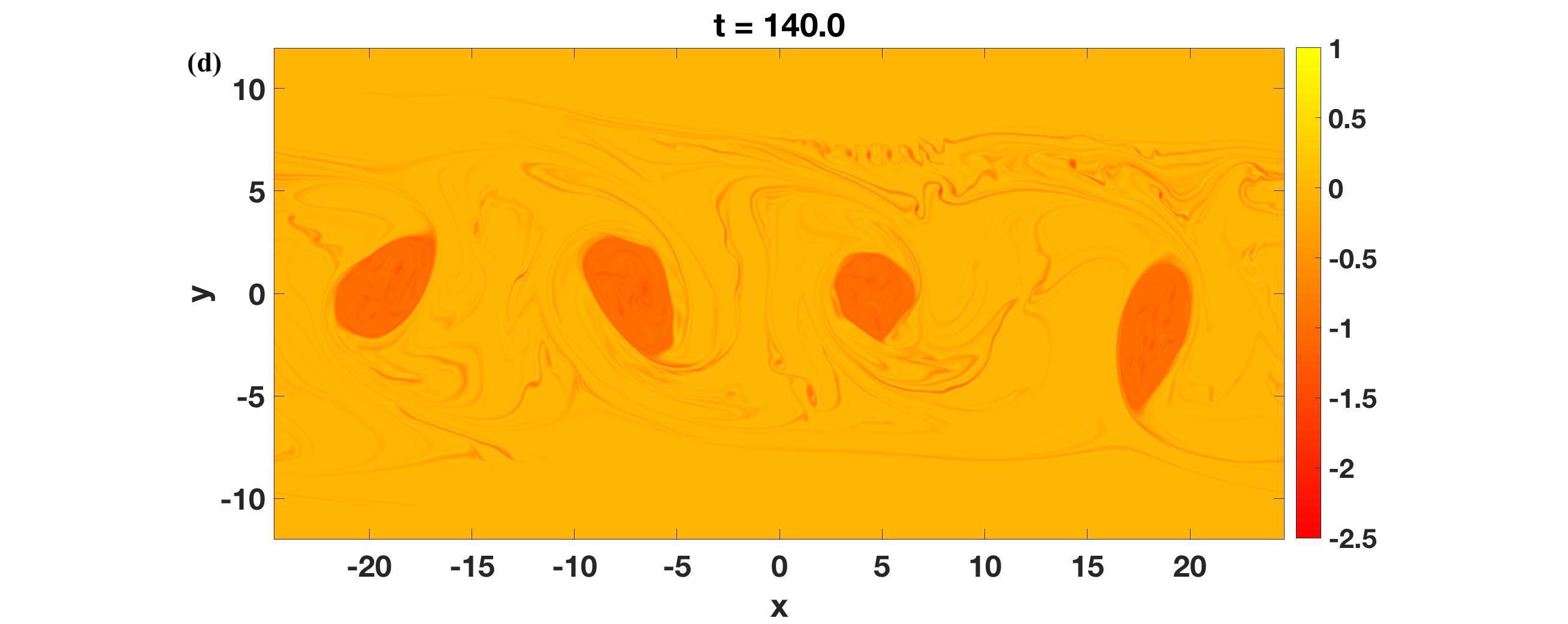}}
\end{tabular}
\caption{Potential vorticity fields for $m=-0.2$ and $L_d=\infty$ at different times (a) $t=10$, (b) $t=30$, (c) $t=50$ and (d) $t=140$.  {{Note that at late times ($t=140$) small scale filaments appear to undergo secondary transition into coherent vortices.}}
}\label{snapshotsm-05LdInfty}
\end{figure}

In figure \ref{snapshotsm-05LdInfty} we add a uniform adverse shear to the Rayleigh profile. In particular the slope of the background shear is settled to $m=-0.2$ while the Rossby radius of deformation is still $L_d=\infty$. The initial disturbance generates again four vortices (b). 
The adverse shear background increases significantly both the number and the resistance of the filamentary structures around the vortices (c). The vortex pairing is weakened by these filaments.
The reduction of the vortex-merging process does hinder the growth of mixing layer indeed. For this reason the vortices size does not grow as much as without adverse shear (i.e. m=0), see fig. \ref{snapshotsm-05LdInfty}d.

\begin{figure}
\begin{tabular}{cc}
\hskip-0.2cm\scalebox{0.095}{\includegraphics[trim={10.5cm 0cm 9.5cm 0cm},clip]{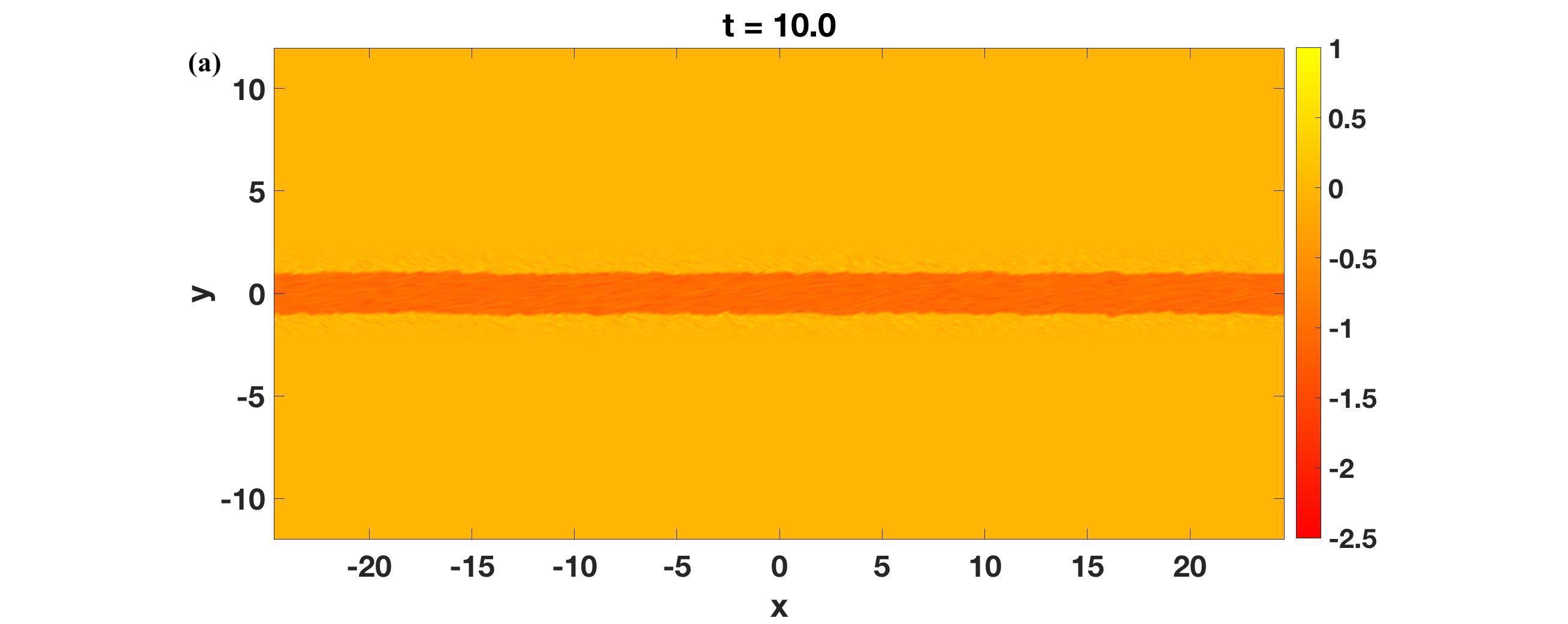}}&\hskip-0.2cm\scalebox{0.095}{\includegraphics[trim={10.5cm 0cm 9.5cm 0cm},clip]{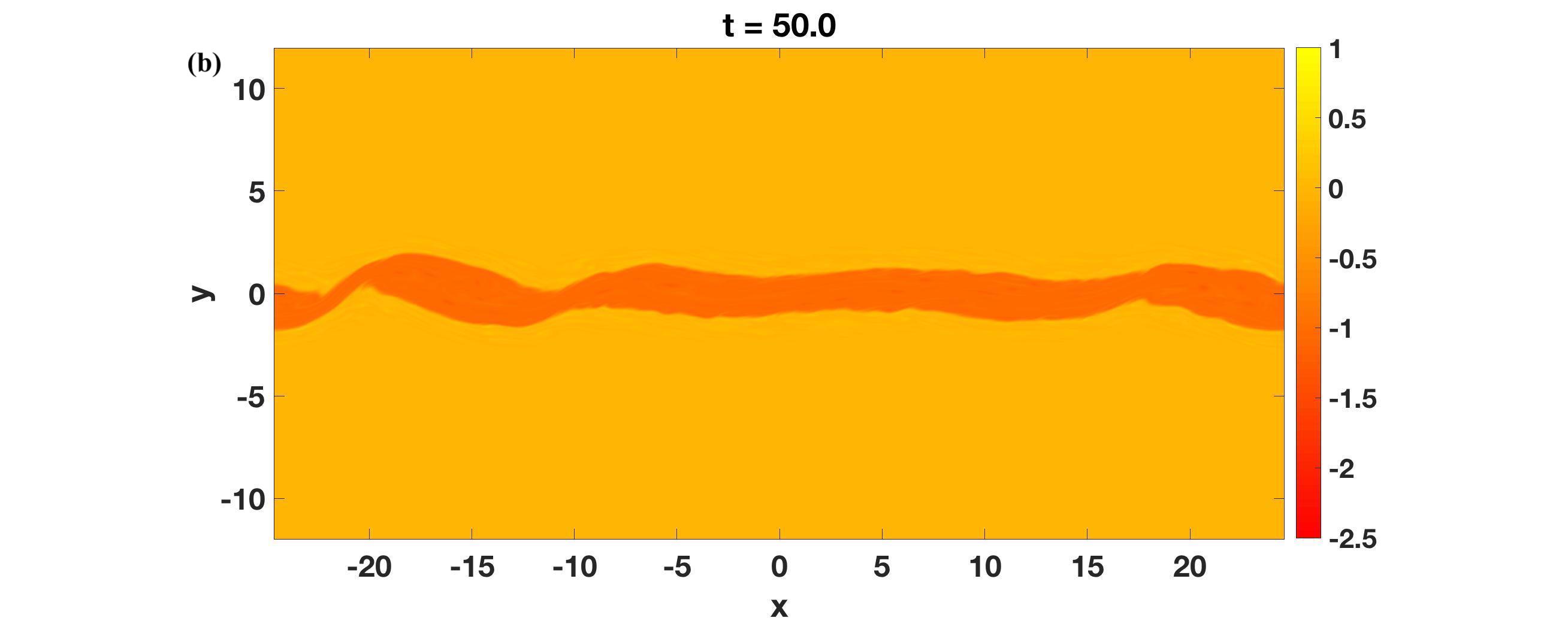}}\\
\hskip-0.2cm\scalebox{0.095}{\includegraphics[trim={10.5cm 0cm 9.5cm 0cm},clip]{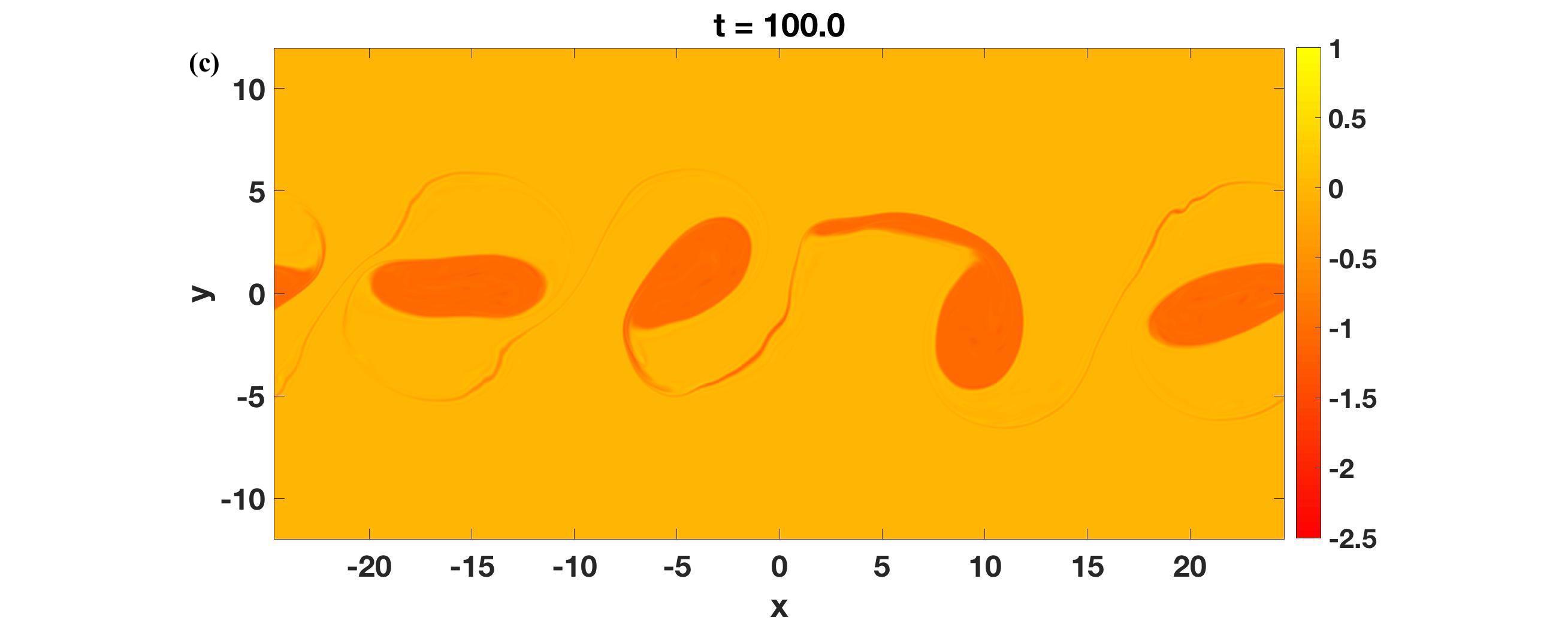}}&\hskip-0.2cm\scalebox{0.095}{\includegraphics[trim={10.5cm 0cm 9.5cm 0cm},clip]{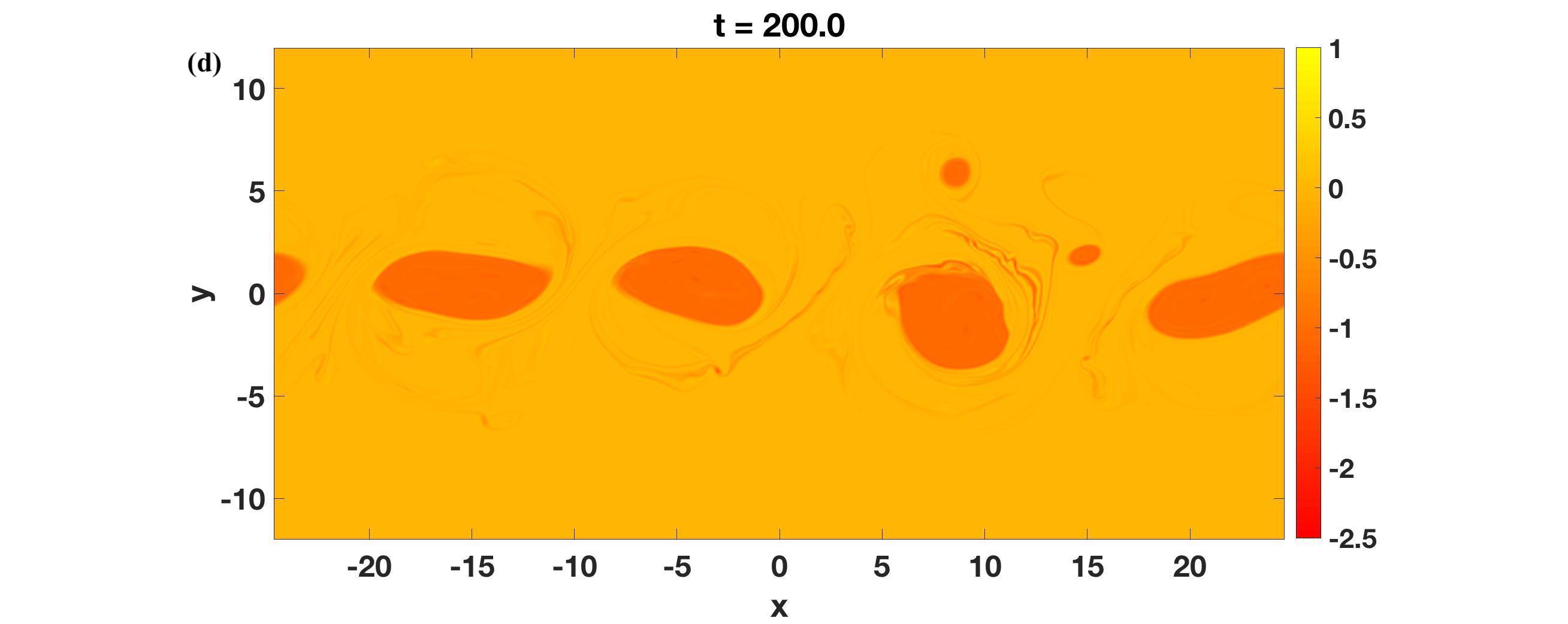}}
\end{tabular}
\caption{Potential vorticity fields for $m=0$ and $L_d=2$ at different times (a) $t=10$, (b) $t=50$, (c) $t=100$ and (d) $t=200$.  { {Similar to the previous figure, torn shreds of vorticity appear to roll into smaller satellite coherent vortices at late times.  
}}
}\label{snapshotsm0Ld2}
\end{figure}

In figure \ref{snapshotsm0Ld2} no uniform shear is added to the Rayleigh profile (i.e. $m=0$) but QG effects are introduced ($L_d=2$). The vortices created by the initial random disturbance are four (b,c). 
In contrast to $m=0$ the vortices do not merge between each other but are slowly rotating (d). { {Since the vortex-merging is absent the mixing layer does not grow as much as when QG effects are neglected.}}

\subsubsection{Momentum Thickness}\label{momthicknesssection}

In this section the spread of the mixing layer is analyzed. Therefore a proper definition for the momentum thickness $\theta$ is needed. We define it as
\begin{equation}
\theta(t)=\int_{-\frac{L^*_y}{2}}^{\frac{L^*_y}{2}}\frac{\bar{u}(y,t)-U_1(y)}{U_2(y)-U_1(y)}\left(1-\frac{\bar{u}(y,t)-U(1)}{U_2(y)-U_1(y)}\right)dy\label{theta}
\end{equation}
where $L_y^*=0.8L_y$ is a vertical length around the shear layer to avoid contamination from the boundaries, $\bar{u}$ is the horizontal velocity averaged along the $x$-direction and $U_1(y)$ and $U_2(y)$) are 
\begin{subequations}
\begin{align}
&U_1(y)=\left\{ \begin{array}{lll} U(y) & {\rm for} &
y>1\\ U(1) &{\rm for}&
y\leq 1
\end{array}\right. \quad {\rm and}
\\
&U_2(y)=\left\{ \begin{array}{lll} U(-1) & {\rm for}& 
y>-1\\ U(y) &{\rm for}&
y\leq -1
\end{array}\right.,
\end{align}
\label{U1U2}
\end{subequations}
respectively. If $m=0$ and $L_d=\infty$ (i.e. the classical Rayleigh profile), $U_1$ and $U_2$ are constant and then eq. \ref{theta} becomes the widely known equation for the momentum thickness \cite{winant1974vortex,biancofiore2014crossover}. 

\begin{figure}
\begin{tabular}{cc}
\hskip-0.1cm\scalebox{0.141}{\includegraphics{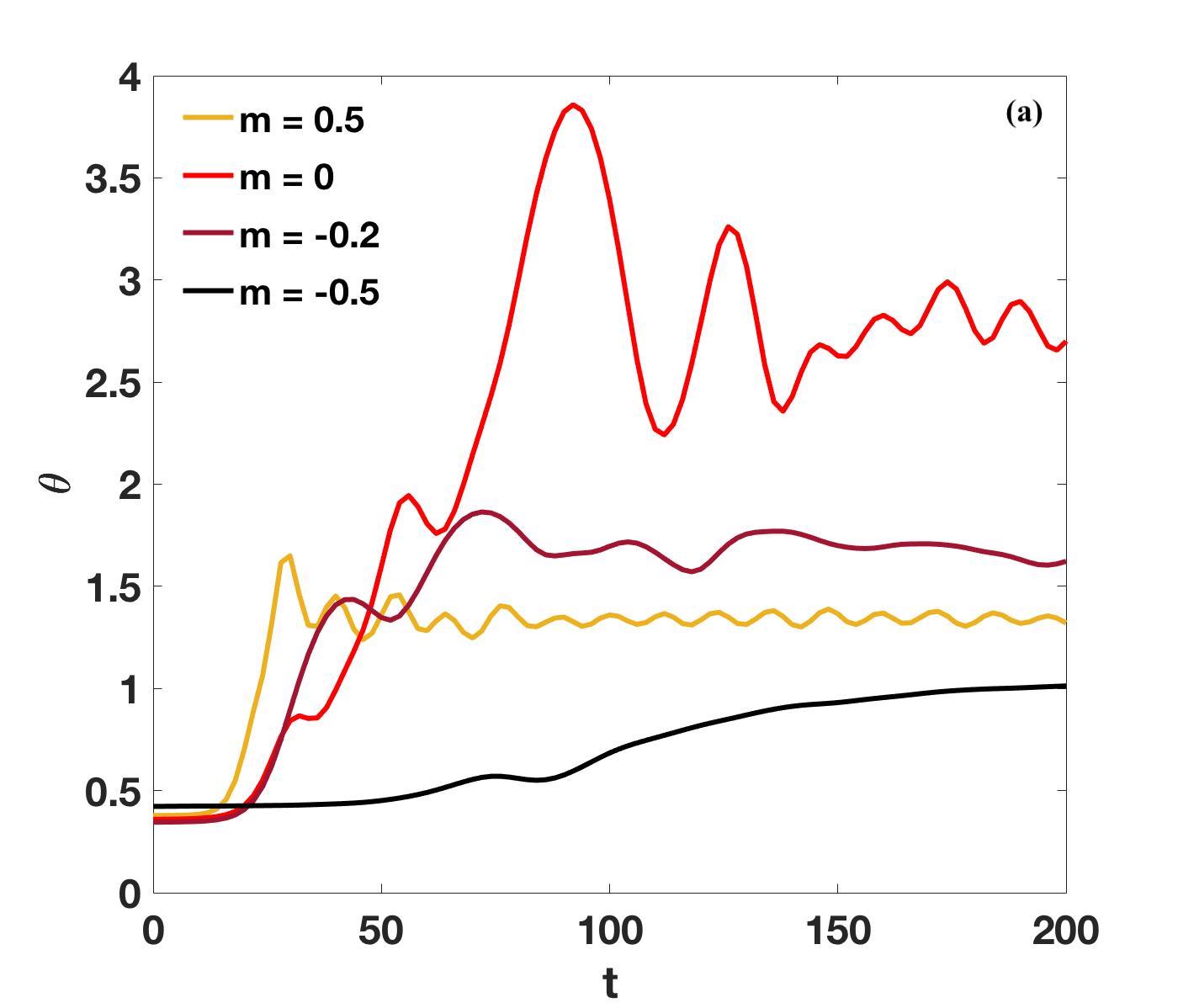}}&\hskip-0.2cm\scalebox{0.141}{\includegraphics{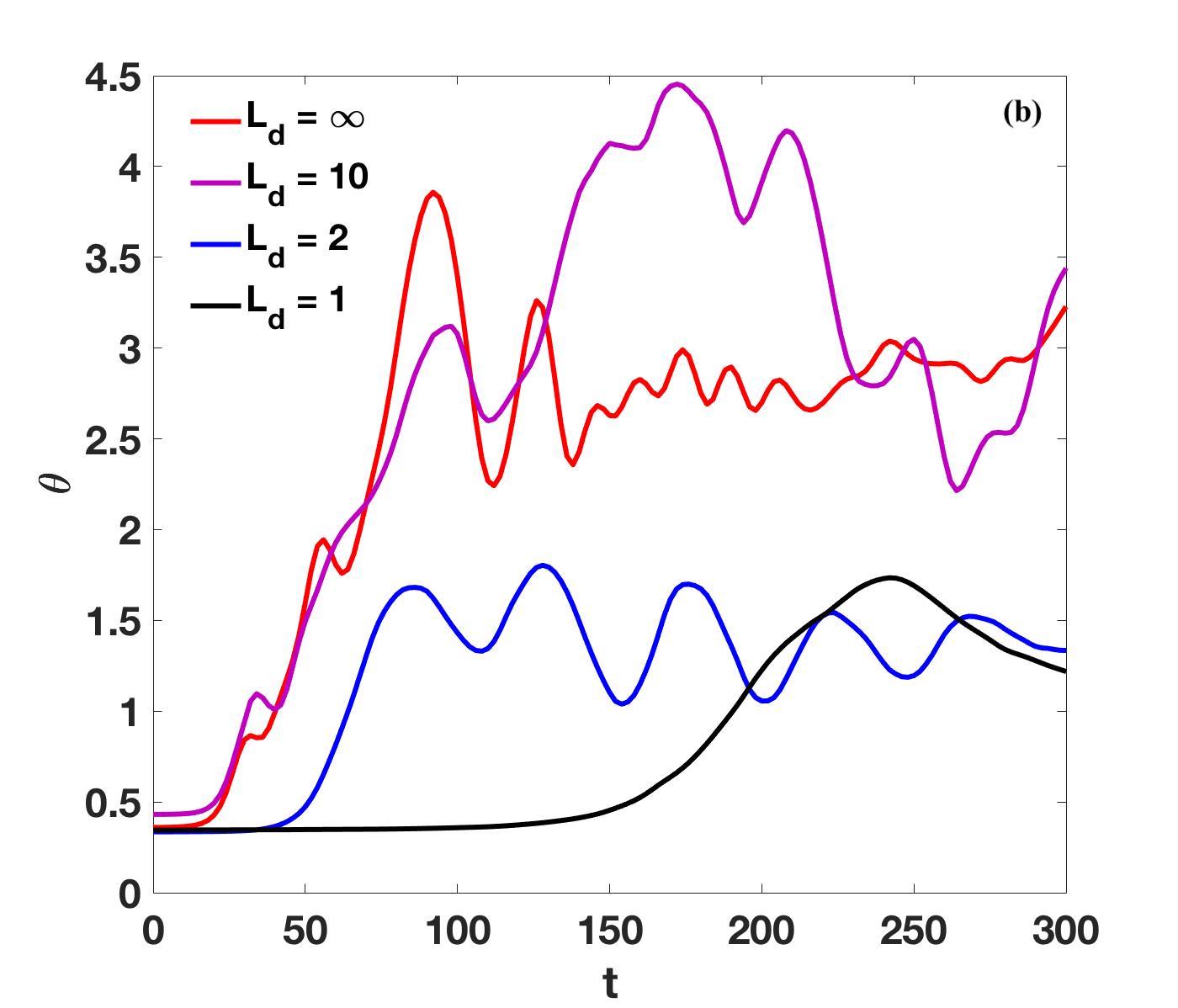}}\\
\end{tabular}
\caption{Temporal evolution of the momentum thickness $\theta$ for (a) $L_d=\infty$ and several values of $m$ and (b) $m=0$ and several values of $L_d$. }\label{momthicknessfigure}
\end{figure}
In figure \ref{momthicknessfigure}a the temporal evolution of the momentum thickness is shown for $L_d=\infty$ and different uniform shear $m$. Interestingly the maximum spread occurs without the background shear (i.e. $m=0$). For negative $m$ we have seen in fig. \ref{snapshotsm-05LdInfty} that the vortex merging is significantly weakened.  This is expected to result in a corresponding weakening of the mixing layer growth. However this dampening is present for positive $m$ also in which the vortex pairing does occur indeed. For this case the momentum thickness cannot significantly diffuse outside the initial shear layer due to the uniform background shear which acts to confine filaments in the shearwise direction by rapidly stretching them out and orienting them into the shearwise direction \cite{umurhan2004hydrodynamic}. Particularly the spreading is hindered by the increased momentum of the sandwiching layers. 

Furthermore the temporal growth of the mixing layer is illustrated in figure \ref{momthicknessfigure}b which introduces QG effects in absence of the background shear. When QG effects are strong the mixing layer spread is damped. This is not surprising since we observed in fig. \ref{snapshotsm0Ld2} that QG effects eliminate the vortex pairing which is the main cause of the growth of mixing layers \cite{winant1974vortex}.

\subsubsection{Perturbation kinetic energy: temporal evolution, spectra}\label{pertkinenergysection}

\begin{figure}
\begin{tabular}{cc}
\hskip-0.1cm\scalebox{0.141}{\includegraphics{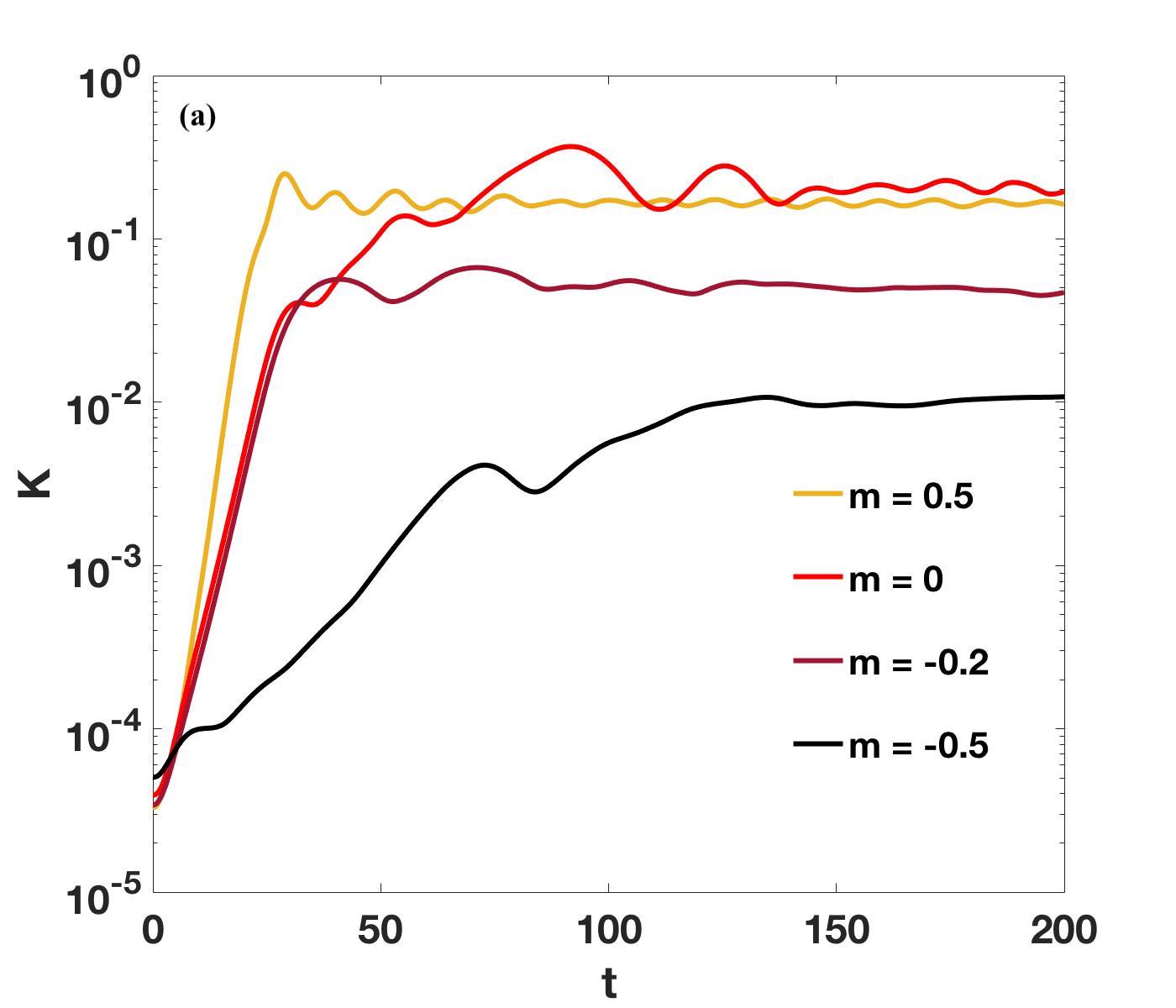}}&\hskip-0.2cm\scalebox{0.141}{\includegraphics{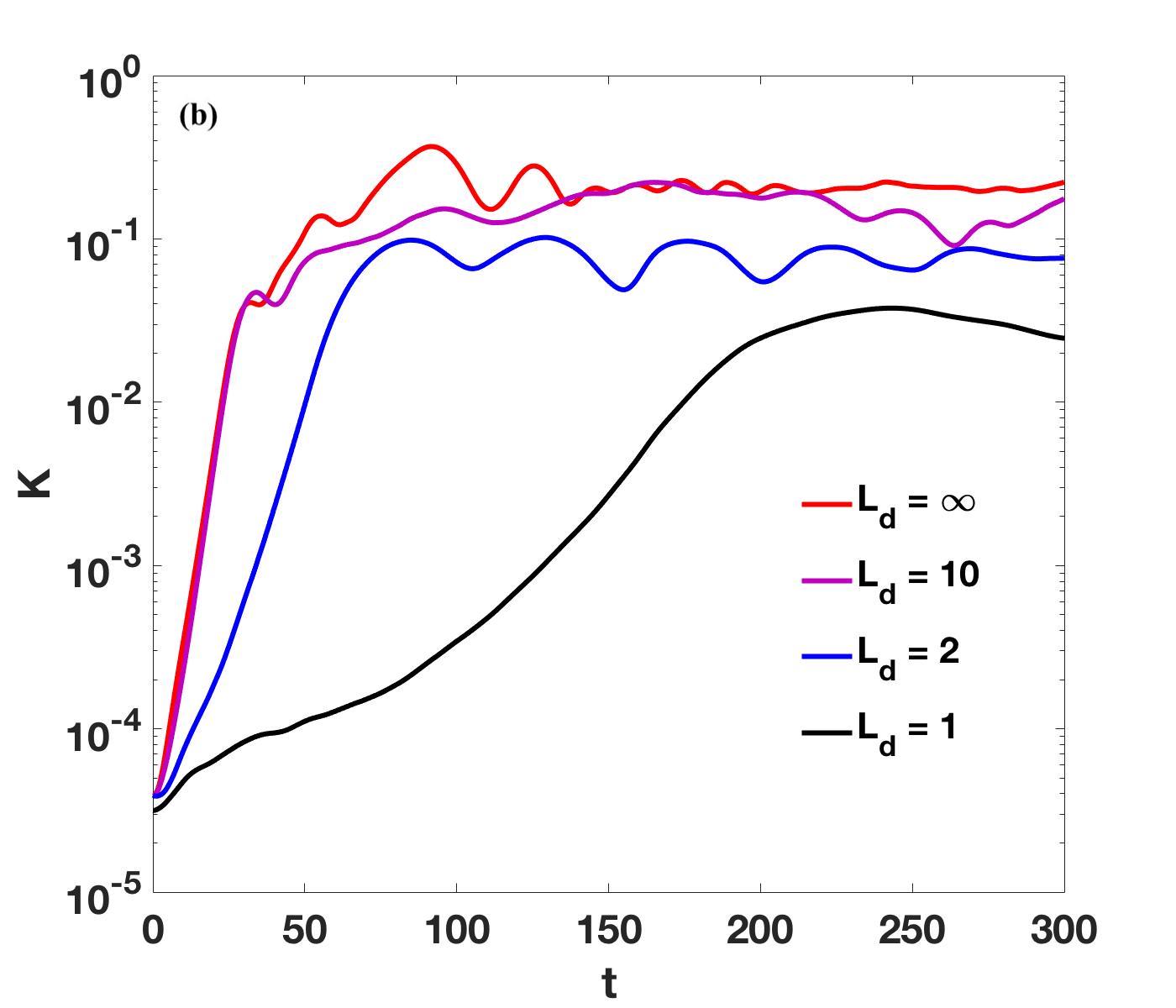}}\\
\end{tabular}
\caption{Spatially averaged perturbation kinetic energy for (a) $L_d=\infty$ and several values of $m$ and (b) $m=0$ and several values of $L_d$. The strength of the turbulence is growing with increasing $m$ and diminishing $L_d$. Note that the domain sizes are $L_x=49.2928$ and $L_y=60$ or $L_y=30$ (when $L_d\leq 2$.}\label{pertkinenergy}
\end{figure}

The temporal evolution of the perturbation kinetic energy averaged along both $x$- and $y$-direction, $K$, is shown in figure \ref{pertkinenergy} for (a) $L_d=\infty$ and different values of $m$ and (b) for $m=0$ and different values of $L_d$. At the beginning the effect of background shear is in agreement with the linear theory as seen in fig. \ref{pertkinenergy}a: if value of $m$ is decreased, the strength of the perturbation increases. However once non-linear effects start to be significant a discrepancy with the linear theory is noticed. While the dampening effect due to an adverse shear is still present, a cooperative shear hinders the growth of $K$ as well once saturation is reached. The additional `boost' in energy to the case without background shear is given by secondary growing modes triggered by the initial unstable mode. This is consistent with the fact that the cut-off number for $m=0$ is larger than $m=0.5$ (see fig. \ref{growthratem}a) with  a wider range of unstable wavenumber.

The non-linear effect of $L_d$ on the temporal evolution of $K$ is in agreement with the linear theory (fig. \ref{pertkinenergy}b). As the Rossby radius of deformation gets smaller the perturbation kinetic energy decreases. The kinetic energy is partially `wasted' along the transversal direction thus there is less available energy to feed the growth of the mixing layer along the vertical direction.

\begin{figure}
\begin{tabular}{cc}
\hskip-0.1cm\scalebox{0.141}{\includegraphics{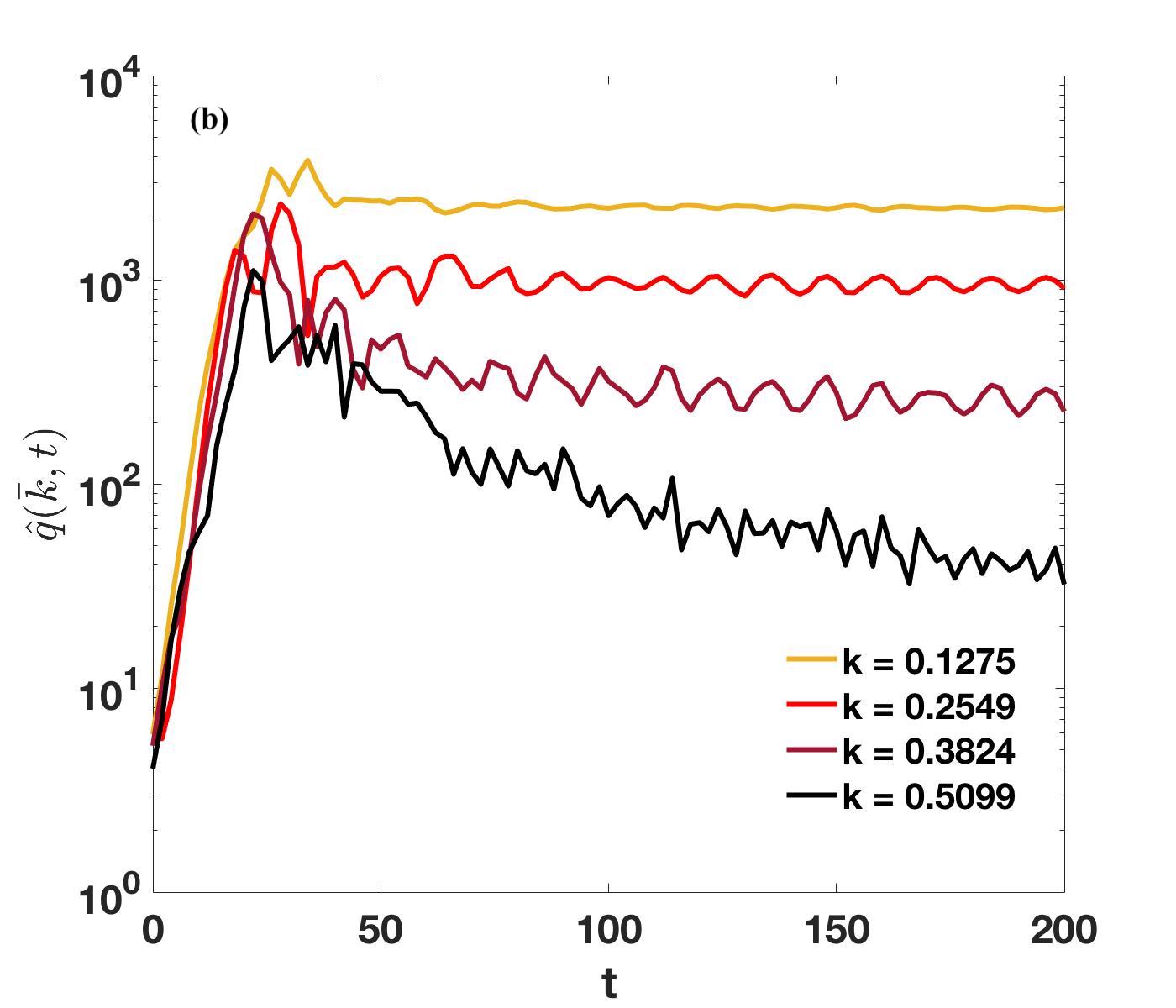}}&\hskip-0.2cm\scalebox{0.141}{\includegraphics{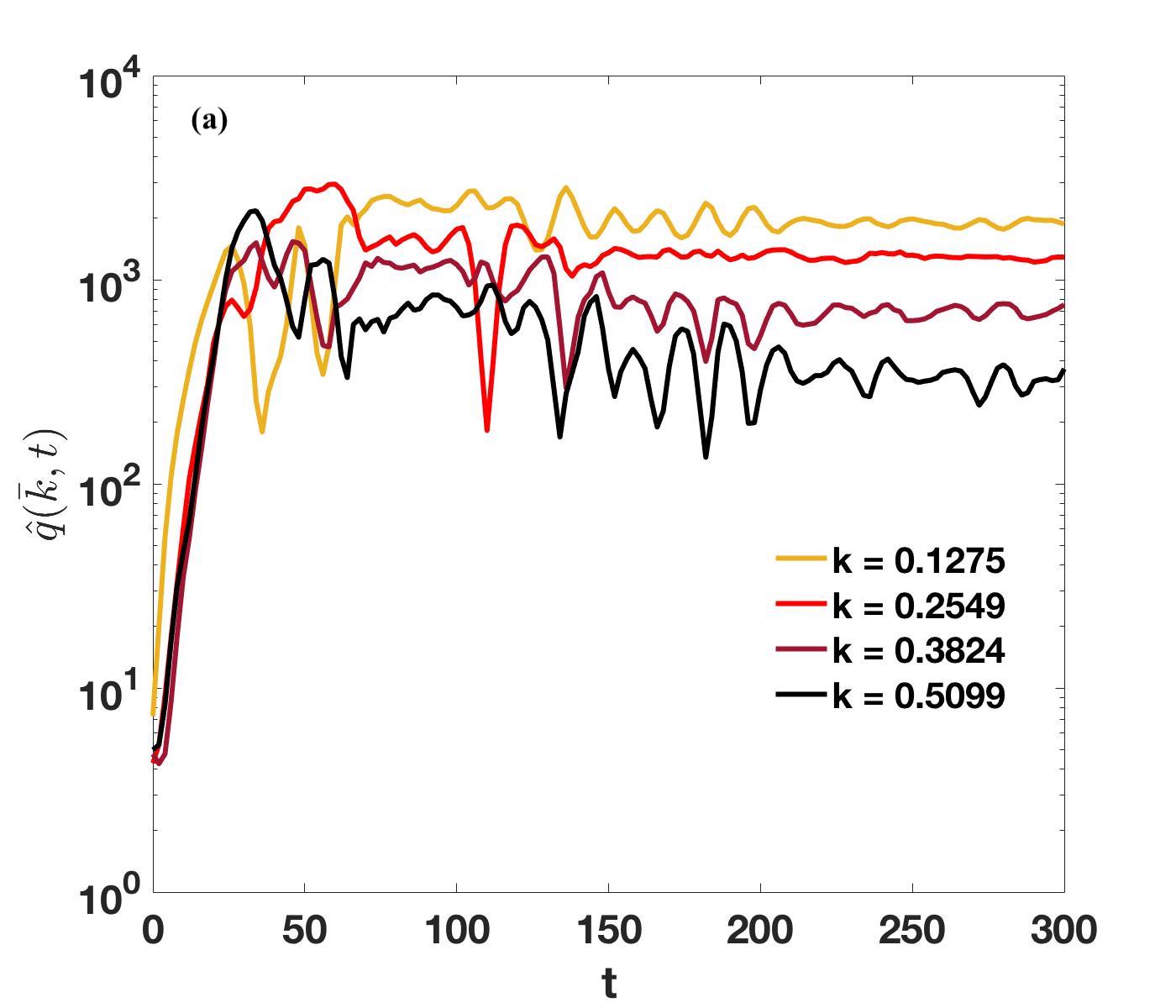}}\\
\end{tabular}
\caption{The Fourier transform of the potential vorticity  $\hat{q}(\bar{k})$ with $k$ equal to $k=0.1275$ (yellow), $k=0.2549$ (red), $k=0.3824$ (dark red), $k=0.5099$ (black) for $L_d=\infty$ (a) $m=0.5$ and (b) $m=0$.}\label{triggeredmodes}
\end{figure}

In figure \ref{triggeredmodes} the $y$-integrated Fourier transform along $x$ of the potential vorticity $\hat{q}(k,t)$ is illustrated for $L_d=\infty$ and (a) $m=0.5$ and (b) $m=0$. In this manner we can observe in detail which wavelengths are triggered. In our simulations we initially trigger a mode with wavelength $\frac{2\pi}{k}$, where $k=0.1275$. For $m=0.5$ this mode dominates both the linear response and the turbulent state. 
For $m=0$ the initial instability triggers modes with different wavenumbers, such as $k=0.2549$ and $k=0.5099$. Then the energy of the final saturated state is not just due to the energy of the initial unstable mode but several modes play a role to reach this value.  { {This overall picture involving only a few mode interactions is consistent with previous studies examining the growth and saturation of mixing layers in porous media using both weakly nonlinear and numerical methods \cite{tamarin2015nonnormal}}}. This confirms our previous hypothesis that the appearance of secondary modes, favored for $m=0$ by the broader range of unstable wavenumbers (see fig. \ref{growthratem}a), enhances the level of turbulent kinetic energy for $m=0$ with respect to $m=0.5$ (see fig. \ref{pertkinenergy}a).

\begin{figure}
\begin{tabular}{cc}
\hskip-0.1cm\scalebox{0.141}{\includegraphics{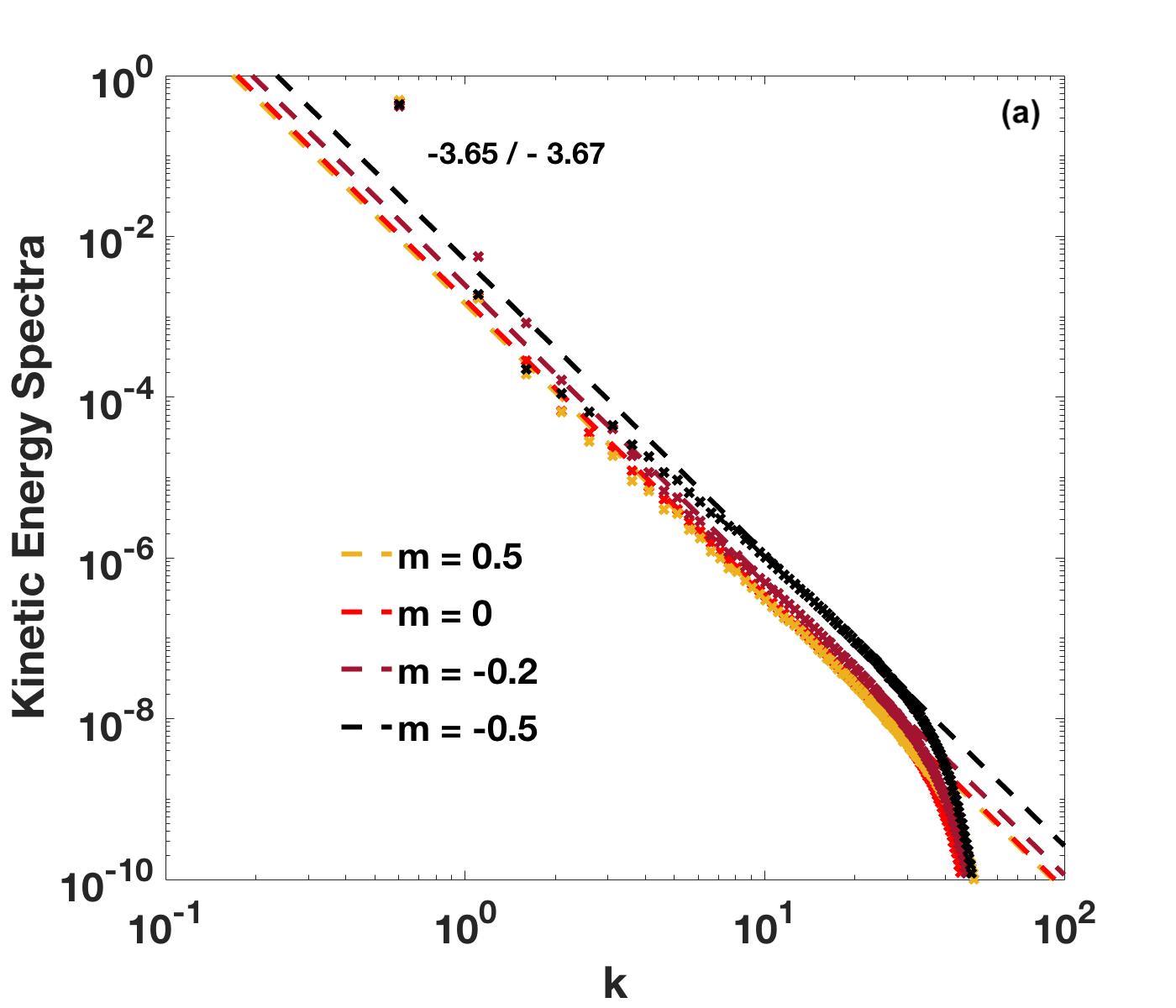}}&\hskip-0.2cm\scalebox{0.141}{\includegraphics{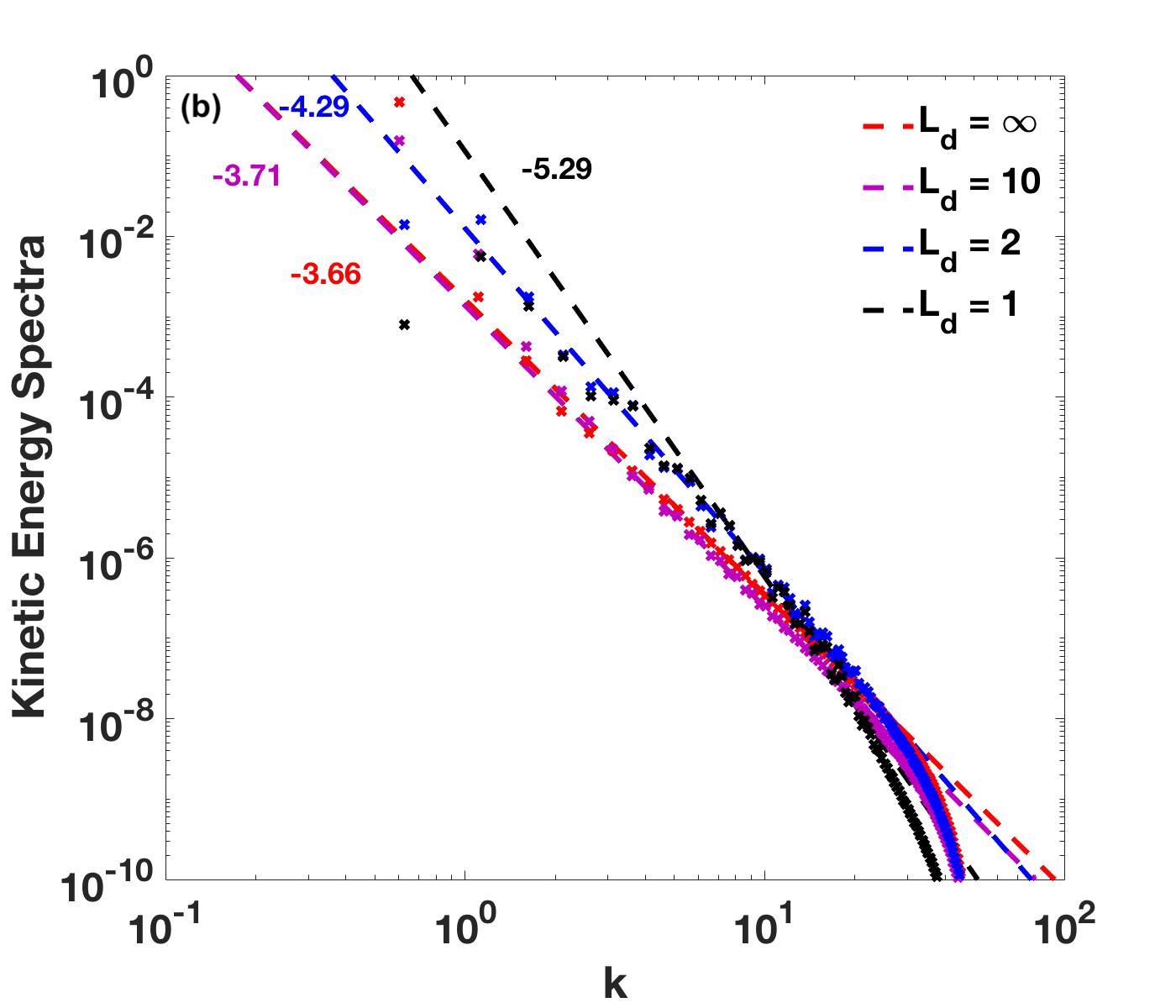}}\\
\end{tabular}
\caption{Time-averaged kinetic energy spectra for (a) $L_d=\infty$ and several values of $m$ and (b) $m=0$ and several values of $L_d$. }\label{spectra}
\end{figure}

The time-averaged spectra of the perturbation kinetic energy $K$ are reported in figure \ref{spectra} for (a) $L_d=\infty$ and several values of $m$ and (b) $m=0$ and several values of $L_d$. Dashed lines depict the slope of the spectra in the inertial range where the enstrophy cascade occurs. 
In both figures the slope is steeper than the  theoretical value of $-3$ predicted by Kraichnan \cite{kraichnan1967inertial} for steady forced-dissipative 2D turbulence. 
In particular for $L_d=\infty$ the slope is between $-3.65$ and $-3.67$ for all the values of $m$. These values are very close to the
$-\frac{11}{3}$ previously reported in other two-dimensional turbulent mixing layer calculations \cite{lesieur1988mixing,biancofiore2014crossover} and, as such,
we view the spectral correspondence, across the whole slew of $L_d = \infty$ models examined here, as further validation 
of our numerical solution methods. {From a kinematic point of view, Gilbert \cite{gilbert1988spiral} has explained that such value is due to the spiral filamentary structures that grow around coherent vortices (see fig. \ref{snapshotsm0LdInfty}f for instance) which are, themselves, characterized by vorticity gradients perpetuating to ever-smaller scales in the spaces between the surviving large-scale coherent structures. (see further discussion below).  Also as evinced figure \ref{spectra}a, we find it significant that the $-\frac{11}{3}$ spectra persists irrespective of the globally imposed constant shear.}



Finally for a finite value of $L_d$ the slope of the spectra significantly increases.
For $L_d=1$ the slope value increases up to $-5.29$. 
Previous authors \cite{basdevant1981study,mcwilliams1984emergence,legras1988high,dritschel2008unifying} found such steep spectra for the enstrophy cascade range (normally between $-4$ to $-6$) due to intermittency of the enstrophy transfer towards small scales.
The enstrophy cascade is fueled by vorticity gradients \cite{weiss1991dynamics,Boffetta_Ecke_2012,regev2016modern}. The significant reduction of filaments around the vortices when $L_d$ is finite (see fig. \ref{snapshotsm-05LdInfty}d) shows how the vorticity gradients are weakened around the vortices. Therefore the direct enstrophy cascade is strongly hindered by QG effects and this is revealed by the steepest spectra shown in fig. \ref{spectra}b. {We discuss this further below in the next section.}


\subsubsection{An Interpretation}\label{Interpretation}

The features examined in the previous sections lead us in adopting the following interpretation inspired, in part, by the picture of self-similar coherent vortex production as propounded in the study of Dritschel {\it et al}\cite{dritschel2008unifying}.
In stationary driven-decaying 2D turbulence calculations at $L_d = \infty$, e.g., as recently reviewed by Boffetta \& Ecke \cite{Boffetta_Ecke_2012}, vortices are produced at some rate at some given length scale.  A direct enstrophy cascade ensues as these coherent vortices undergo structural breakdown that generate filaments that become elongated, bent and sheared, and eventually, fully populate the spaces between the (quasi) coherent structures of the injection scales.  Ultimately this process cascades down to the dissipation scales giving rise to the nearly $k^{-3}$ energy spectra predicted by Kraichnan\cite{kraichnan1967inertial} to characterize these flows. 

Of course, current best resolved numerical simulations \cite{Boffetta_Musacchio_2010} show that the spectrum takes on the form $k^{-3-\delta}$ with $\delta \approx 0.65$.  Boffetta \& Musacchio's study \cite{Boffetta_Musacchio_2010} involved conducting a series of numerical experiments with increasing resolution and showed that $\delta$ progressively gets smaller as resolution and Reynolds number increase. They conclude that this system ought to asymptotically yield $\delta \rightarrow 0$ as $Re \rightarrow \infty$.

The ability to shred apart vortices is directly a function of the relative distance between (say) two vortices {\it and} their mutual ability to influence/advect their edges which is, in turn, a function of $L_d$ -- larger values of $L_d$ implies vortex structures can induce flow at longer range.  Obviously, for a given separation of two given coherent vortices of given size, filament production should be preferentially promoted
in models where $L_d$ is largest. With this physical view, it therefore seems reasonable to suppose that spatially dense filaments with strong spatial gradients become rarer in models where $L_d$ decreases.

\par

{ We view the features of the mixing layer within this same framework, but we read its implications in the following way: The mixing layer has no steady production rate of vortices but has, instead, a short duration production of vortices owing to the roll-up/breakdown of the original layer itself.  This generates filaments which wind up in response to the velocity field of the aggregate vorticity distribution and, when applicable, the background imposed uniform vorticity.  In cases when several coherent vortices are born from the primary breakdown,  copious filamentary structure is produced as in all of the $L_d = \infty$ cases we examined (e.g., see figures \ref{snapshotsm0LdInfty}-\ref{snapshotsm-05LdInfty}). Within these subset systems, coherent vortices tend to merge when the background shear is zero which, in turn, further produces more filamentary structures both inside and outside the resulting merged structure -- this is especially evident in the latter stages of development as shown in figure \ref{snapshotsm0LdInfty}d.  In both cases, filaments are seen to also roll-up and generate smaller scale coherent structures as well -- suggestive that a self-similar process is at play.   In this sense, it appears that the filaments are not nearly as space filling as they are in the classical stationary forced-dissipated 2D turbulence systems -- with its usual $k^{-3}$ direct enstrophy cascade spectra -- since coherent vortices appear to be produced alongside spiraling filaments at ever decreasing spatial scales \cite{Santangelo_etal_1989}.  While these coherent structures remain stable to destruction, vorticity stays locked within them choking-off the production of more filaments at smaller scales \cite{Tabeling_2002}.  It is important to note that a flow that only yields a self-similar distribution of vortices at all scales down to the dissipation scales will exhibit an energy spectra of the form $k^{-5}$, see Refs. \onlinecite{dritschel_1995,dritschel2008unifying}.
We can think of this idealized self-similar vortex system as one end-member state, with the other end-member being steady forced-dissipated 2D turbulence with its space-filling spiraling vortex filaments.
In this sense, and very much in line with Gilbert's thinking \cite{gilbert1988spiral}, we rationalize the $k^{-11/3}$ spectral slope of these mixing layer experiments to be a reflection of the fact 
that they, as a setting, lie in between these two end-member states since
both self-similar coherent structures are present alongside ever-tightly spiraling vortex filaments.
}
\par
{   The simulations with $L_d$ small like those shown in figure \ref{snapshotsm0Ld2}, may be interpreted in a similar vein. These mixing layers also undergo breakdown but their coherent child PV structures are relatively stable.  This stability is due entirely to the fact that individual coherent structures have more limited dynamical reach compared to their larger $L_d$ counterparts.  That is to say, the flow induced by a coherent PV structure is weaker and greatly restricted in range compared to a similar structure with a larger value of $L_d$.  In these smaller $L_d$ settings, unless individual coherent PV structures get very close one another, vortex-vortex mergers cannot occur as readily as they do in $L_d\rightarrow \infty$ conditions.  With merging events becoming infrequent, fewer filaments get produced, and most of the PV remains locked within these larger scale structures.  This gives rise to the steepness of the energy spectra observed in these cases -- which increasingly resembles the idealized self-similar vortex end-member case with its characteristic $k^{-5}$ energy spectrum.}

\section{Conclusions}\label{conclusions}

In this work we have revisited by means of the KW perspective the linear and non-linear stability of quasi-geostrophic vortex filaments under the influence of a background shear.
 These filaments were modeled by a family of quasi-Rayleigh profiles subject to a uniform shear $my$. 
Through the KW perspective the instability of the quasi-Rayleigh profiles is viewed as the interaction of two counter-propagating Rossby waves created at the two PV edges (i.e. where the PV is discontinuous). 

We confirm that an adverse (favorable) shear stabilizes (destabilizes) the flow, as seen by Dritschel \cite{dritschel1989stabilization}. Owing to the KW framework we observed that the optimal phase locking configuration is shifted towards smaller (larger) wavenumbers if $m$ is positive (negative), while the interaction between the two CRWs is totally independent on the background shear. Phase locking occurs in a range of wavenumbers in which the interaction is more (less) favored for positive (negative) $m$ than when the background shear is absent. 
The introduction of QG effects has a stabilizing influence on the Rayleigh profile independent of the presence of a background shear. As the Rossby deformation radius $L_d$ decreases the interaction between the two CRWs is weakened. This agrees with the common understanding that QG effects diminish the reach of the edge waves. 

We have examined the non-linear breakdown of a vortex filament in the nominally $\infty$-Reynolds number limit
by using a superviscosity operator to drain power that builds up on individual grid scales due to turbulent cascade through the inertial regime.
 In the simulations, the quasi-Rayleigh profile is again used to model the vortex filaments while the background shear is viewed as immutable and a continuous source of external forcing. 
We have validated our numerical code and model set-up by comparing our linear predictions to our numerical simulations during the linear growth phase.  
 We have demonstrated the usefulness of the KW perspective through this equivalence (section \ref{NLS}).  The KW viewpoint says that at the two edges of the mixing-layer, only the PV jump and the mean flow evaluated at these edges determines the stability of the mixing-layer.  Indeed,  the mean streamwise velocity fields in the background shear enhanced quasi-Rayleigh profile we have adopted and the profile analyzed in the linear analysis are very different. However the two flow fields share the same two core characteristics, i.e., the PV jumps across strip edges and the total mean flow at these edges are equivalent in both profiles. Both profiles are shown to yield the same growth rates.
 \par
 
With regards to the quality of the mixing-layer's development and breakdown, we find several properties: A negative background shear hinders the vortex pairing and then the growth of the mixing layer while with a positive background shear the growth mechanism is still obstructed but this is due to the increased momentum of the external layers. A finite value of the Rossby radius of deformation also hinders the vortex pairing then the mixing layer growth is maximized with a configuration with an infinite $L_d$. While QG effects have a similar influence on linear and non-linear analyses, we observe discrepancies between these analyses concerning the impact of the background shear on the turbulence strength. Particularly, they differ for a positive $m$ since in the non-linear simulations the most energetic configuration is without a background shear while for the linear analysis the gravest growth rate was for $m=0.5$. For zero background shear additional modes are subharmonically activated by the disturbance while this does not occur for $m=0.5$ since the range of unstable wavelengths is narrower. 

When QG effects are absent, the slope of the enstrophy cascade in the kinetic energy spectra is very close to the value of $-\frac{11}{3}$, typical of 2D turbulent mixing layers \cite{lesieur1988mixing,biancofiore2014crossover}.  This value for the slope is intermediate between $-3$ (predicted by Kraichnan \cite{kraichnan1967inertial} for forced-dissipated 2D turbulence systems) and $-5$ (found for a self-similar distribution of vortices at all scales \cite{Santangelo_etal_1989,dritschel_1995,dritschel2008unifying}) showing how 2D turbulent mixing layers lie between these two end-member states. When a finite $L_d$ is introduced the spectra become steeper.
Mixing layer coherent structures have a more limited dynamical reach as $L_d$ diminishes that hinders the merging process. Rarer vortex-vortex mergers strongly decrease the filaments production around the vortices. The consequent absence of filaments obstructs the enstrophy cascade steepening thus the spectra.

Our work promotes a mechanistic resonance-based perspective toward understanding how vorticity waves behave in a QG environment. To our knowledge, this is the first study to examine QG shear instability {\emph{through the KW perspective lens}}. Expanding to other geophysical or astrophysical flow systems a similar study can be conducted to evaluate the influence of QG effects on interacting gravity waves \cite{rabinovich2011vorticity} or Alfv\`en waves \cite{heifetz2015interacting}. It would be interesting to observe if and to what degree these wave interactions become weakened in a QG environment, similarly to what occurs for CRWs.

Moreover revisiting QG effects can be seen as a reasonable and intuitively useful model
framework toward better understanding
 three-dimensional effects from the vantage of the KW perspective. From our study of the QG model, the third dimension may be `simply' seen as a direction into which the kinetic energy of CRWs may be stored as potential energy in the form of fluid thickness variations.  The richness of the possible modes of action are, thus, already apparent from our analysis of the QG model and their interpretations. Many new types of instabilities are known to become manifest with the introduction of the (spanwise) direction, i.e., normal to the flow. It would be intriguing to adapt the KW perspective to capture three-dimensional instabilities in more realistic and accurate models. This kind of study could have impact toward rationalizing not only geophysical flows but also more applied flow systems connected to industrial and technological applications \cite{biancofiore2012counterpropagating,biancofiore2015interaction}.

\bibliographystyle{unsrt}
\bibliography{AdverseShear_PRF.bib}
\end{document}